\documentclass[useAMS,usenatbib]{mn2e}
\usepackage{epsfig,lscape} 
\usepackage{natbib}
\usepackage{amsmath}

\newcommand{\ha}{\hbox{H$\alpha$}}
\newcommand{\hb}{\hbox{H$\beta$}}
\newcommand{\oii}{\hbox{[O\,{\sc ii}]}}
\newcommand{\oiii}{\hbox{[O\,{\sc iii}]}}

\newcommand{\nii}{\hbox{[N\,{\sc ii}]}}

\title[The mass-metallicity relations for gas and stars in star-forming galaxies]{The mass-metallicity relations for gas and stars in star-forming galaxies: strong outflow vs variable IMF}
\author[J. Lian et al.]
{Jianhui Lian$^{1}$\thanks{ljhhw@mail.ustc.edu.cn (J. Lian); daniel.thomas@port.ac.uk (D. Thomas)},
Daniel, Thomas$^{1}$, Claudia Maraston$^{1}$, Daniel Goddard$^{1}$,
\newauthor Johan Comparat$^{2}$, Violeta Gonzalez-Perez$^{1}$, Paolo Ventura$^{3}$\\
$^{1}$Institute of Cosmology and Gravitation, University of Portsmouth, Burnaby Road, Portsmouth, UK, PO1 3FX\\
$^{2}$Max-Planck-Institut f\"{u}r extraterrestrische Physik, Postfach 1312, Garching, Germany, 85741\\
$^{3}$INAF–Osservatorio Astronomico di Roma, via Frascati 33, I-00077 Monteporzio, Italy
}

\begin{document}
\maketitle

\begin{abstract}
We investigate the {mass-metallicity relations for the gaseous (MZR$_{\rm gas}$) and stellar components (MZR$_{\rm star}$)} of local star-forming galaxies based on a representative sample from SDSS DR12.
The mass-weighted average stellar metallicities are systematically lower than the gas metallicities. 
This difference in metallicity increases toward {galaxies with lower masses and reaches 0.4-0.8 dex at 
10$^9M_{\odot}$ (depending on the gas metallicity calibration). As a result,} the {MZR$_{\rm star}$} is much steeper than the {MZR$_{\rm gas}$}. The much lower metallicities in stars compared to the gas in low mass galaxies implies dramatic metallicity evolution with suppressed metal enrichment at early times. 
The aim of this paper is to explain the observed large difference in gas and stellar metallicity and to infer the origin of the {mass-metallicity} relations. 
To this end we develop a galactic chemical evolution model accounting for star formation, gas inflow and outflow.
By combining {the observed mass-metallicity relation for both gas and stellar components}
to constrain the models, we find that only two scenarios are able to reproduce the observations. Either strong metal outflow or a steep IMF slope at early epochs of galaxy evolution is needed. Based on these two scenarios, for the first time we successfully reproduce the observed {MZR$_{\rm gas}$ and MZR$_{\rm star}$} simultaneously, together with other independent observational constraints in {the} local universe. 
Our model also naturally reproduces the flattening of the {MZR$_{\rm gas}$} at the high mass end leaving the  {MZR$_{\rm star}$} intact, as seen in observational data.
\end{abstract}


\section{Introduction}
Metallicity is a key parameter for understanding galaxy evolution as it plays an important role in many fundamental galactic physical processes, such as star formation, gas cooling and collapse, stellar evolution, and dust formation. 
To measure the metal abundance of various astrophysical objects, different techniques are proposed. A widely-used metallicity proxy is oxygen abundance measured from emission lines in star-forming regions of galaxies. Since oxygen is the most abundant heavy element, the oxygen abundance is a good approximation to the total metal abundance.
For non-star-forming regions without emission lines in the spectra, absorption lines and stellar population models are employed. 
Generally, the metallicity obtained from emission lines refers to the gas component, while the metallicity determined from stellar continuum and absorption line fitting refers to the stellar component.

A correlation between the stellar mass (or luminosity) and gas metallicity of star-forming galaxies was found decades ago \citep{lequeux1979} according to which massive galaxies have higher gas metallicities. 
Later studies (e.g., \citealt{tremonti2004,kewley2008}) have confirmed this relation with large galaxy samples from the Sloan Digital Sky Survey (SDSS; \citealt{york2000}). Besides the dependence on stellar mass, other studies have found further dependences of gas metallicity on other physical properties at a given mass, such as specific star formation rate (sSFR, \citealt{ellison2008}), star formation rate (SFR, \citealt{mannucci2010,lopez2010,andrews2013}), and stellar age \citep{lian2015}. These higher dimensional relations could provide additional constraints to the processes that regulate the metal enrichment in galaxies. In addition to gas metallicity, also the stellar metallicity of galaxies is found to correlate with the stellar mass \citep[e.g.][]{gallazzi2005,panter2008,thomas2010}, suggesting the {mass-metallicity relation (MZR)} already existed at early epochs of galaxy evolution \citep{savaglio2005,erb2006, maiolino2008,yuan2013}. 

The {MZR} is one of the most important global scaling relations of galaxies that sets strong constraints on models of galaxy formation and evolution. Although {the mass-metallicity relation for gas (MZR$_{\rm gas}$) and stars (MZR$_{\rm star}$) have benn}
observationally established for a long time, their physical {drivers are} still under debate. 
Many explanations have been proposed. One explanation for the MZR$_{\rm gas}$ is metal-enriched gas outflow induced by supernova explosions \citep{larson1974,tremonti2004,kobayashi2007,scannapieco2008}. Galaxies of lower masses are expected to be more efficient in removing the freshly enriched gas owing to the shallower gravitational potential well. 
Alternatively, a dilution effect caused by metal-poor gas {inflow} could also potentially decrease metallicities at the low mass end of 
the {MZR$_{\rm gas}$} \citep{dalcanton2004,dave2010}, assuming a longer  time scale for the {inflow} in lower mass galaxies. Another possibility is a mass-dependent star formation efficiency (SFE; \citealt{brooks2007,calura2009}). 
In this scenario, more massive galaxies are more efficient in converting their gas reservoirs into stars which results in lower gas-to-stellar mass ratios and higher metal abundance. Based on a chemical evolution model, \citet{calura2009} found that the {MZR$_{\rm gas}$} could be naturally reproduced by a lower SFE in less-massive galaxies. 
Lastly, variations in the initial mass function (IMF) are proposed as another possible factor that could play a role in establishing the 
{MZR$_{\rm gas}$} \citep{koppen2007}. This plethora of different explanations with various mass-dependent galaxy properties for the 
{MZR$_{\rm gas}$} suggests the presence of degeneracies between the parameters that regulate metal enrichment in the interstellar medium (ISM) of star-forming galaxies. 

Most studies on chemical enrichment in galaxies focus on the gas and stellar metallicities separately. However, the simultaneous study of both sets stringent constraints on galaxy chemical evolution. The stellar metallicity carries information about the early epochs of chemical enrichment in galaxies, while gas metallicity reflects more recent evolutionary processes. The degeneracy between the driving parameters of the {MZR} can therefore be broken by studying gas and stellar metallicity simultaneously. In general, average stellar metallicities in galaxies are found to be lower than gas metallicities at all masses, with the largest discrepancy at low mass \citep{gonzalez2014}. Recent attempts to explain this discrepancy have failed. Using the semi-analytic model L-GALAXIES, combined with a state-of-the-art galactic chemical evolution model, \citet{yates2012} were able to reproduce the {MZR$_{\rm gas}$} of local star-forming galaxies but predicted a stellar metallicity significantly higher than the observation. 
\citet{rossi2017} found similar results based on a high-resolution cosmological hydrodynamical simulation of EAGLE. Clearly, the challenge is to reproduce the relatively low average stellar metallicity at a given gas metallicity. 

Prompted by this challenge, we construct a simple galactic chemical evolution model. 
To better understand the role played by the various processes in the metal enrichment history of galaxies, we are looking for scenarios that  
reproduce the {MZR$_{\rm gas}$ and the MZR$_{\rm star}$} simultaneously. The paper is {organized} as follows. In \textsection 2 we introduce the sample selection and determination of gas and stellar metallicity in galaxies from the SDSS. Then we explain the ingredients of the chemical evolution model and extensively explore its parameter space in \textsection 3. In \textsection 4 we show the finely-tuned models that match the observation. We also give a discussion in \textsection 5 and a final summary in \textsection 6.  
Throughout this paper, we adopt the cosmological parameters with $H_0=70\, {\rm km s^{-1} Mpc}^{-1}$, $\Omega_{\Lambda}=0.70$ 
and $\Omega_{\rm m}=0.30$.

\section{Observational data}
\subsection{Sample selection}
We select a sample of local star-forming galaxies from the SDSS
Data Release 12 \footnote{http://www.sdss.org/dr12/} (DR12; \citealt{alam2015}) with the following criteria:
(1) reliable stellar mass measurements and $M_*>10^9 {\rm M}_{\odot}$; 
(2) redshift range of $0.02\leq z\leq 0.05$ to ensure mass completeness;
(3) high specific star formation rate, ${\rm log(sSFR)}>-0.6*{\rm log}(M_*/{\rm M_{\odot}})-4.9$, in order to select star-forming galaxies;
(4) signal-noise-ratio (SNR) of strong emission lines (including \hb,\oiii$\lambda\lambda4959,5007$,\ha,\nii$\lambda6584$) above 5; 
(5) not classified as composite or AGN objects by the demarcation of \citet{kewley2001} in the BPT diagram \citep{baldwin1981}. 
To quantify the robustness of the mass estimate, we use the difference between the 84th and 16th percentiles of the stellar mass estimate as the error and require the error to be less than 0.3 dex. 
{Galaxies with stellar mass below $10^9{\rm M}_{\odot}$ usually show relatively low quality in the continuum of their SDSS spectra. It is more difficult to reliably derive the stellar metallicity from the spectra continuum with low SNR. The mass cut at $10^9$ is empirical but the result will not change significantly if we adopt a different mass cut.}
The final sample consists of 4633 galaxies. 

Figure 1 shows the galaxy distribution in the mass versus sSFR diagram (left-hand panel) and mass versus $u-i$ colour diagram (right-hand panel). 
Black data points represent  
SDSS galaxies with $0.02<z<0.05$ and $M_*>10^9{\rm M_{\odot}}$, while the blue points are the galaxies selected here. 
The dashed line indicates the sSFR criteria for star-forming galaxies {which is designed to best separate the two main galaxy populations based on visual inspection.}
It can be seen that most of the selected galaxies occupy the blue cloud star-forming region in the mass-colour diagram, suggesting the sSFR cut for selecting star-forming galaxies is robust. 
{We also test a constant sSFR cut and find our conclusions are independent on the details of this selection criteria.}
The estimates of stellar mass and SFR are taken from 
the MPA-JHU value-added catalogue \citep{kauffmann2003,brinchmann2004} {named `galSpecExtra' in SDSS DR12}. A Kroupa IMF \citep{kroupa2001} is assumed in deriving the mass and SFR. 
We use the emission line {measurements from the catalogue named `galSpecLine' in SDSS DR12 which are derived by
MPA-JHU spectroscopic reanalysis described in \citet{tremonti2004} and \citet{brinchmann2004}.  
Stellar absorption has been taken into account in measuring the emission line fluxes by fitting the spectra continuum with a stellar population model.}
We correct for the galactic internal extinction using the Balmer decrement method and adopting the Milky Way extinction law \citep{ccm89}. 

\begin{figure*}
\centering
\includegraphics[width=15cm]{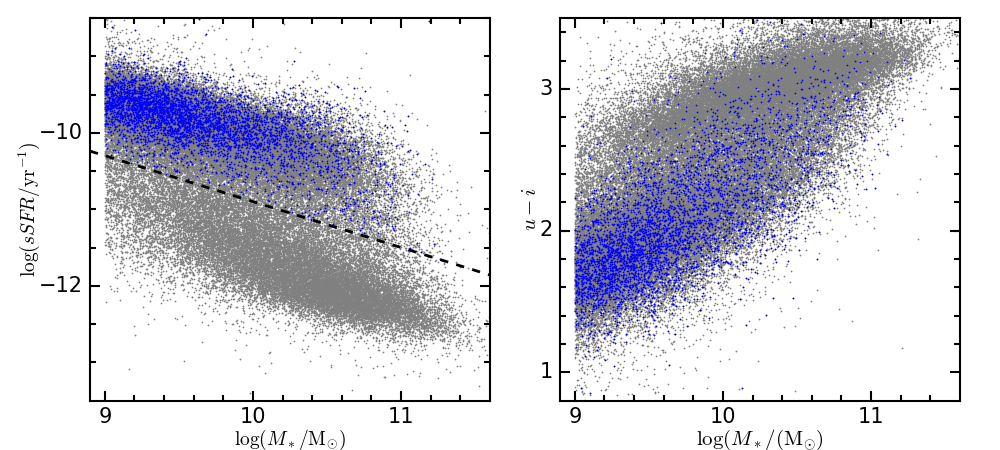}
\caption{Galaxy distribution in mass versus sSFR (left-hand panel) and mass versus $u-i$ colour diagram (right-hand panel). Black data points represent  
SDSS galaxies with $0.02<z<0.05$ and $M_*>10^9{\rm M_{\odot}}$ and blue points are the galaxies selected here. The dashed line indicates the sSFR criteria used to select star-forming galaxies. 
}
\label{figure1}
\end{figure*}

\subsection{Metallicity determination}
\subsubsection{Gas metallicity}
There are many methods proposed to determine the gas metallicity of emission-line galaxies \citep{kewley2008,maiolino2008}.
Among them, the so-called `Te' method is considered to be one of the most reliable methods. 
To obtain the electron temperature, two excitation lines of the same ion are needed. 
The most widely-used pair of excitation lines are \oiii$\lambda$4363 and \oiii$\lambda$5007. 
However, the \oiii$\lambda$4363 line is usually weak, hence a number of empirical methods calibrated to the Te method have been proposed.
These empirical methods use a single or a combination of strong emission line ratios for the calibration.
Some of the most-widely used ones are the `N2 method' (\nii$\lambda$6584/\ha; \citealt{pp04}), the `O3N2 method' ((\oiii$\lambda5007$/\hb)/(\nii$\lambda$6584/\ha);
\citealt{pp04}), and the `R23 method' ((\oii$\lambda3727$+\oiii$\lambda\lambda$4959,5007)/\hb; \citealt{pilyugin2005}).
Rather than calibrating to the metallicity determined by the Te method, some other studies use similar 
emission line ratios but calibrate to the theoretical {photoionization} model \citep{m91,kewley2002,kk04}. 
Large discrepancies of up to 0.7~dex have been found between the derived metallicities by different calibrations \citep{shi2005,kewley2008}. 
The physical origin of this discrepancy is still not fully understood. 
In this work, considering the large discrepancy in different gas metallicity calibrations, we adopt the {empirical} `N2' method calibrated by 
\citet{pp04} and the theoretical `R23 method' calibrated by \citet{kk04} to derive the gas metallicity. 
{The empirical N2 method is only valid for $-2.5<{\rm log(\nii\lambda6584/\ha)}<-0.3$.
We have also tested another empirical gas metallicity calibrator `O3N2' from \citet{pp04}, and most of our results derived from the N2 method are also valid when using the O3N2 method.}
It is worth noting that, according to the comparison of these gas metallicity calibrations
in \citet{kewley2008}, the empirical N2 calibration gives almost the lowest estimate of gas metallicity while the theoretical R23 method gives the highest estimate. 

Figure~2 shows the distribution of star-forming galaxies selected in \textsection2.1 in the {mass versus gas metallicity} diagram. 
The {MZR$_{\rm gas}$} from the empirical N2 and the theoretical R23 methods are shown in cyan and blue colours, respectively.  
The {light shaded regions indicate the 16th to 84th percentiles of the distributions while the dark shaded regions represent the error of the median value at a given mass.}
To compare with the literature, we include the {MZR$_{\rm gas}$} of local star-forming galaxies compiled by \citet{kewley2008} which use
the empirical O3N2 methods from \citet{pp04} and the theoretical method from \citet{tremonti2004}.
It can be seen that the empirical methods give a much lower metallicity than the theoretical methods. The empirical N2 method and theoretical R23 method give the lowest and highest metallicity values, respectively. 
We can also see that the trend of more massive galaxies having higher gas metallicity is independent of the metallicity determination method. It is interesting to note that
the {MZR$_{\rm gas}$} flattens at $M_*>10^{10}{\rm M}_{\odot}$.
This is true for all metallicity determination methods (see Figure 2 in \citealt{kewley2008}).

\begin{figure}
\centering
\includegraphics[width=\columnwidth,viewport=15 10 650 620,clip]{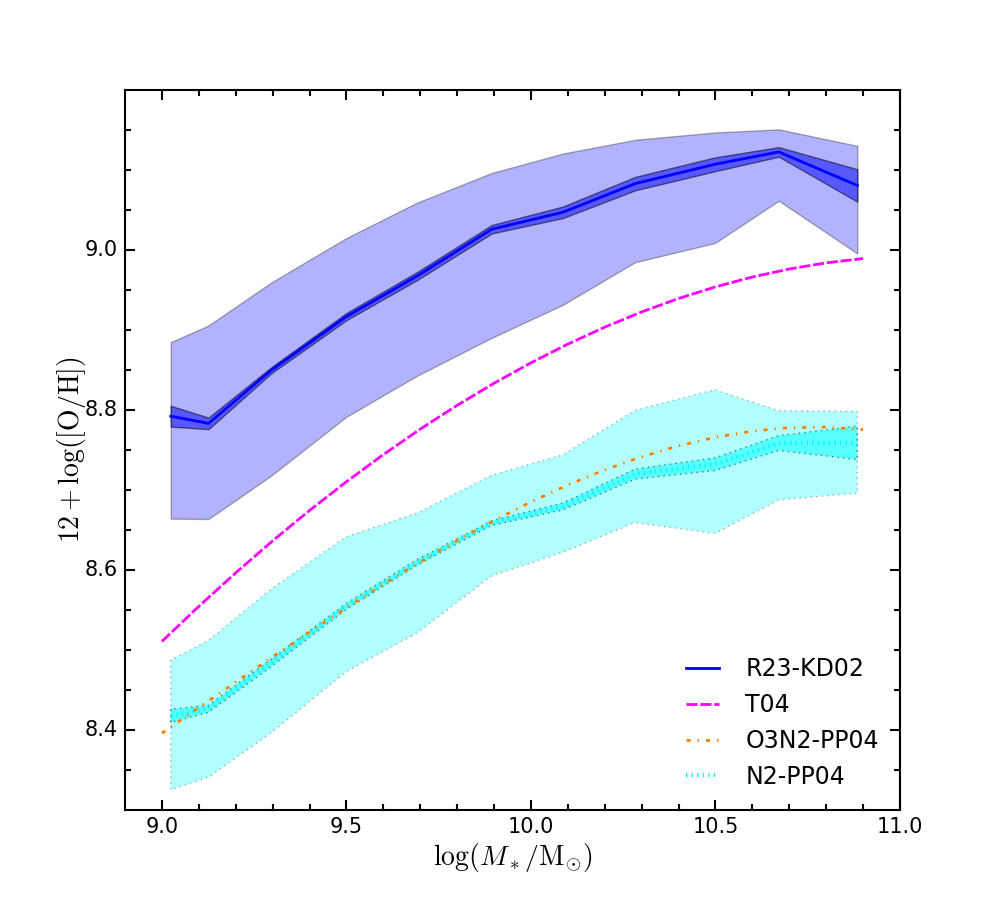}
\caption{{Mass-metallicity} relation {for the gaseous component} of local star-forming galaxies with gas metallicity determined by various methods. 
The green dotted line represents the {MZR$_{\rm gas}$} from the empirical N2 method by \citet{pp04}, while the blue solid line indicates the relation from the theoretical R23 method by \citet{kk04}.
Light shaded regions indicate the 16th to 84th percentiles of the distributions while dark shaded regions represent the error of median metallicity at a given mass.
The same relations in the literature compiled by \citet{kewley2008} are included
with gas metallicity calibrations
from \citet{tremonti2004} (T04) and \citet{pp04} (O3N2-PP04). }
\label{figure2}
\end{figure}

\subsubsection{Stellar metallicity}

Stellar metallicity is an important observable as it carries information about the early chemical enrichment history of galaxies and provides additional constraints on the chemical evolution model. 
Also the stellar metallicity of a galaxy is a difficult measurement to make, because of the well known degeneracy between age, metallicity and dust when {analyzing} galaxy spectra. Yet, modern advancements in stellar population {modeling} and improvements in the quality of observational data have led to significant progress in the analysis of stellar population parameters, allowing robust measurements of metallicity to be obtained for large statistical samples. In this work, we use the full spectral fitting code FIREFLY (\citealt{wilkinson2015,wilkinson2017}; Goddard et al. in prep) and the stellar population models of Maraston and Str\"omb\"ack \citep{maraston2011}, with a Kroupa IMF \citep{kroupa2001} to derive stellar metallicity. FIREFLY uses a $\chi^2$ {minimization} technique to fit Single Stellar Population (SSP) models to an input galaxy spectrum. The code uses an iterative algorithm to combine arbitrarily weighted linear combinations of SSPs, in order to find the best fit model given the data and employs minimal priors, allowing maximal exploration of the parameter space. This has been shown to be a good way to accurately recover the properties of galaxies. The velocity dispersions needed as input to FIREFLY are adopted from the {MPA-JHU} catalogue, which are in good agreement with the measurements by \citet{thomas2013}. We adopt the {\em mass-weighted} mean stellar metallicity as obtained from FIREFLY. The other outputs of FIREFLY, such as stellar mass, are also consistent with literature \citep{wilkinson2015}.

Figure~3 shows the {MZR$_{\rm star}$} of our sample and comparison with the literature. Similar to Figure~2, the light red shaded region indicates the 16th to 84th percentile of {\em mass-weighted} stellar metallicity while the dark red shaded region represent the error of median stellar metallicity in each mass bin. 
The green dashed line represents the {\em light-weighted} {MZR$_{\rm star}$} for comparison. Light-weighted metallicities are generally {higher} than mass-weighted metallicities, as young populations tend to be more metal-rich in the composite populations derived by FIREFLY (see also Goddard et al. in prep). 
For comparison, we also include the {MZR$_{\rm star}$} of star-forming galaxies raised by \citet{peng2015} where the metallicity measurements were taken from \citet{gallazzi2005}. {It can be seen that the {MZR$_{\rm star}$} derived by FIREFLY is in good agreement with the relation in \citet{peng2015}.}
In a related study we present the direct comparison with literature metallicity measurements for the full galaxy population and find a good agreement of the {MZR$_{\rm star}$} using FIREFLY (Goddard et al. in prep).

Similar to the gas metallicity, we find a positive correlation between stellar metallicity and galaxy mass, with
more massive galaxies having higher stellar metallicity on average, albeit some considerable scatter as shown in Figure 3. A similar {MZR$_{\rm star}$} has been found for local galaxy population including passive galaxies \citep{gallazzi2005,panter2008,thomas2010,johansson2012} using early released SDSS data. Relevant for the present study, it turns out that the mass-weighted {MZR$_{\rm star}$} is much steeper than the {MZR$_{\rm gas}$}. Since light-weighted stellar metallicity reflects the last stellar generations, it is flatter with mass and closer to the gas metallicity compared to the stellar-weighted stellar metallicity. 
This agrees with the finding by \citet{gonzalez2014} from CALIFA IFU survey that the metallicities of the younger populations are higher and match the gas metallicity. Since the chemical evolution model predicts the intrinsic and not light-weighted metallicity, we shall use the mass-weighted determinations. 

\begin{figure}
\centering
\includegraphics[width=\columnwidth,viewport=20 10 650 620,clip]{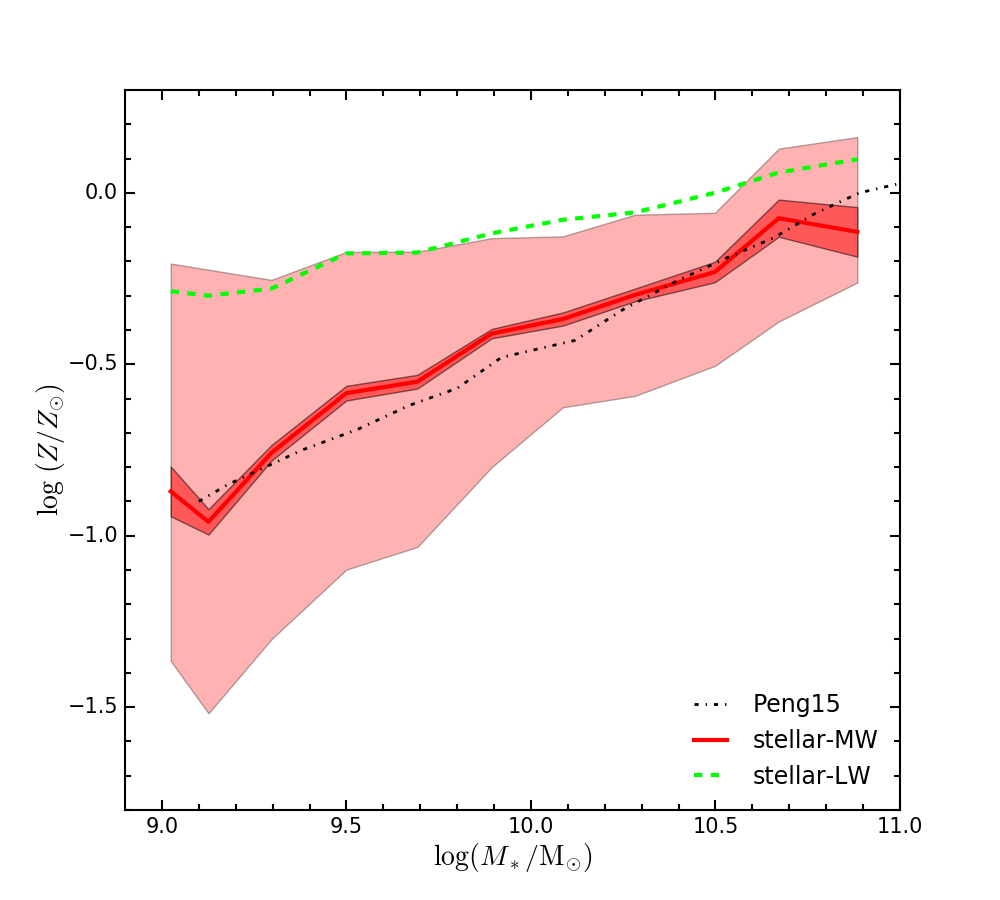}
\caption{
{Mass-metallicity relation for the stellar component} of local star-forming galaxies selected as detailed in \textsection2.1. Similar to Figure 2, the solid line and light shaded region indicate the median and 16th to 84th percentile of the mass-weighted stellar metallicities, respectively. The dark shaded region shows the $1\sigma$ error of median stellar metallicity in each mass bin. The green dashed line represents the light-weighted {MZR$_{\rm star}$}. Black dash-dotted line denotes the {MZR$_{\rm star}$} of star-forming galaxies derived by \citet{peng2015}.} 
\label{figure3}
\end{figure}

\subsection{Gas and stellar metallicity comparison}
\begin{figure*}
\centering
\includegraphics[width=15cm]{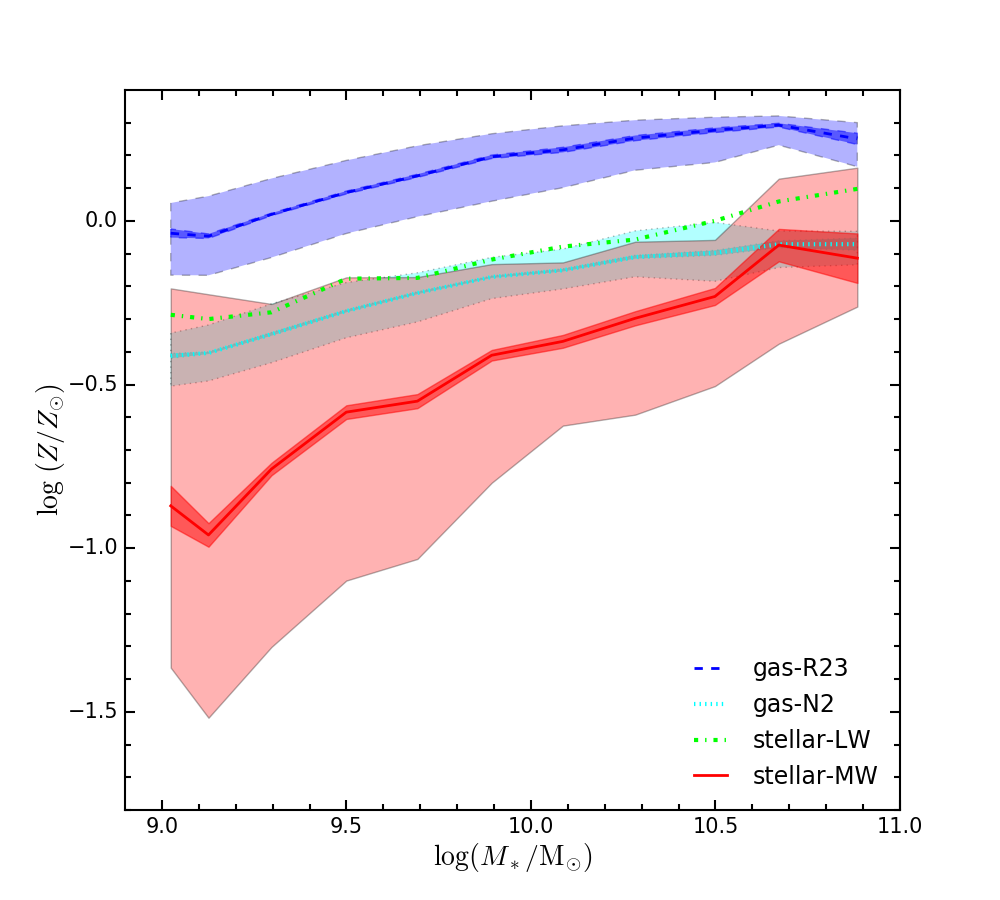}
\caption{Comparison between the gas and stellar metallicity of local star-forming galaxies (see \textsection2.1 for details on the selection). Gas metallicities derived by the N2 and R23 methods are shown as cyan and blue colour, respectively. The mass-weighted {MZR$_{\rm star}$} is shown in red colour. The light shaded regions indicate the 16 and 84 percentiles {while dark shaded regions represent the error of median metallicity.}
}
\label{figure4}
\end{figure*}

In Figure 4 we directly compare the gas metallicity measured by the N2 and R23 methods with the mass-weighted stellar metallicity. 
To enable a direct comparison between gas and stellar metallicity, we {normalize} the gas oxygen abundance to solar oxygen abundance. 
Since the stellar metallicity derived by FIREFLY is calibrated based on a solar metal abundance of 0.02, for a fair comparison, we adopt the corresponding solar oxygen abundance of $12+\log({\rm O}/{\rm H})=8.83$ \citep{anders1989}. Adopting the updated solar metallicity and oxygen abundance from \citet{asplund2009} will enhance both the gas and stellar metallicity by $\sim 0.14$ dex. Therefore the relative difference between the gas and stellar metallicity, from which our main result is derived, is not significantly affected by the absolute value of solar abundance. 
 
{It should be noted that the gas and stellar metallicity are both measured from the SDSS fiber spectra which typically covers the central region of nearby galaxies. 
The two metallicity measurements are comparable since they reflect the properties of the same region of galaxies, but they may not represent the integrated properties of galaxies. As star-forming galaxies usually show a negative gradient in both gas and stellar metallicity (Lian et al. prep), correcting the aperture effect will end with sightly lower gas and stellar metallicity measurements. Since the stellar metallicity gradient is typically steeper in more massive galaxies, the correction effect will be more significant in stellar metallicity of massive star-forming galaxies. As a result, the aperture-corrected MZR$_{\rm star}$ is expected to be flatter than the MZR$_{\rm star}$ directly derived from fiber spectra. To quantitatively determine the aperture effect and study the spatial distribution of gas and stellar metallicity of galaxies, one need to use integral field spectroscopy data which is the subject of our next paper in preparation.}

From the comparison we can see that star-forming galaxies usually have
higher metallicities in the ISM compared to what is locked in stars, which is consistent with the result from CALIFA  \citep{gonzalez2014}. Also, it is interesting to note that this difference is mass dependent, changing from 0.8~dex (0.4~dex) at $10^9 M_{\odot}$ to 0.4~dex (0~dex) at $10^{11} M_{\odot}$ based on gas metallicities derived by the R23 ({N2}) method. 
In other words, the {MZR$_{\rm star}$} of star-forming galaxies is much steeper than the {MZR$_{\rm gas}$}.
{The gas metallicity derived through the N2 method is systematically lower than the light-weighted stellar metallicity by $\sim0.1$ dex. Since the metal enrichment history of galaxies is generally monotonous, it is natural to find a higher metallicity in the ISM than in the stars. Therefore, this opposite pattern found here when adopting the N2 method to derive gas metallicity suggests that the N2 method possibly underestimates the intrinsic metallicity in gas of star-forming galaxies.}
 
Generally, the stellar metallicity carries information about the early epochs of chemical enrichment in galaxies while the gas metallicity reflects more recent evolutionary processes. 
Therefore, a much higher metallicity in the ISM than in stars implies a dramatic metallicity evolution with extremely low metallicity at the early evolutionary epochs of galaxies, i.e.\ metal enrichment at early epochs must be substantially suppressed. 
The mass dependence of the difference between gas and stellar metallicity suggests that
this metallicity evolution is {also mass-dependent and} strongest in the lowest mass galaxies. 

The trend of gas metallicity with stellar mass has been reproduced by semi-analytical galaxy evolution models 
\citep[e.g.,][]{calura2009,yates2012} and hydrodynamical simulations \citep{guo2016,rossi2017}. However, the {MZR$_{\rm star}$} predicted by these models does not match the observed relation. The models generally overestimate stellar metallicities, and predict a flatter slope for the {MZR$_{\rm star}$} than is observed. In the next section, we will explore possible physical scenarios that can explain the gas and mass-weighted stellar metallicity in star-forming galaxies using a full galactic chemical evolution model. 

\section{Chemical evolution model}  

\subsection{Model ingredients}
\subsubsection{Gas content}
The chemical evolution of a galaxy is closely related to the evolution of its gas content, which is driven
by many fundamental processes, including gas accretion from the halo, gas outflow through feedback, and galaxy internal star formation activity. 
The time-evolution of the gas content in a galaxy can be described by the following basic equation \citep{tinsley1980}:
\begin{equation} \label{eq1}
\dot{M}_{\rm g} = -\psi+\dot{M}_{\rm ej,g}+\dot{M}_{\rm inf}-\dot{M}_{\rm out}.
\end{equation}
Here $\psi$ is the star formation rate and $\dot{M}_{\rm ej,g}$ denotes the mass ejection from intermediate and massive stars back  
into the ISM. The last two variables represent the inflow and outflow rates. 

In the present study we adopt a single inflow process with an inflow rate declining exponentially with cosmic time.
Two parameters are needed to describe this 
inflow rate; the initial strength $A_{\rm inf}$ and the declining time scale $\tau_{\rm inf}$. 
To {parameterize} the galactic outflow in the model, we define a quantity $f_{\rm out}$ that represents the mass fraction of stellar ejecta that are expelled from the galaxy. No interaction of the outflow with the ISM in the galaxy is assumed. 
Very high resolution simulations show that heated particles are enriched \citep{maclow1999,creasey2015},
this result qualitatively agrees with the metal-enriched outflows assumed in our models.

The SFR of a galaxy was found to be proportional to its gas content leading to a universal star formation law. 
A widely applied star formation law is the `Kennicutt-Schmidt (KS) law' calibrated by \citet{kennicutt1998} 
between the surface densities of the SFR and the gas as 
$\sum_{\rm SFR}=A_{\rm ks}\times(\sum_{\rm gas}/M_{\odot}{\rm pc}^{-2})^{n_{\rm ks}} M_{\odot}{\rm yr}^{-1}{\rm kpc}^{-2}$.
$A_{\rm ks}$ represents the KS law coefficient which is found to be $2.5*10^{-4}$ in \citet{kennicutt1998}.
For the sake of simplicity, the parameter $A_{\rm ks}$ in our model is scaled to the original value of $2.5*10^{-4}$. 
The parameter $n_{\rm ks}$ denotes the exponent of the KS law which is 1.4 in \citet{kennicutt1998}. 
To better understand the role played by the star formation law in the chemical evolution model, 
we allow both the coefficient and power law index to vary in the model. 
We adopt a constant effective radius of 5~kpc for galaxies of different masses in order to calculate the surface density from integrated properties. 
Adopting a different size will affect the derived surface density of gas and thus the SFE which could also be tuned by the initial inflow strength, coefficient of KS law. Therefore, we fix the size of galaxies in the model to minimize the number of model parameters. 
With the description of gas inflow, gas outflow and star formation rate, we are able to predict the final galaxy stellar mass for a given initial inflow strength. The star formation history of the model galaxy then follows from the above equation. Hence, the final stellar mass of a model galaxy is mainly determined by the initial inflow strength $A_{\rm inf}$ and further regulated by the inflow timescale $\tau_{\rm inf}$. Higher galaxy masses are generally obtained through higher values for the parameter $A_{\rm inf}$.

\subsubsection{Metal abundance}
Non-primordial elements are {synthesized} during the life cycle of stars and ejected at the late stages of stellar evolution through stellar winds, planetary nebulae (PN), or supernova (SN), depending on the mass of the star on the main sequence. 
Similar to the gas content, the equation for the time evolution of each metal element, after applying our definition of outflow, can be written as \citep{tinsley1980}
\begin{equation} \label{eq2}
\dot{M}_{\rm i} = -\psi*Z_{\rm i}+\dot{M}_{\rm ej,i}+\dot{M}_{\rm inf}*Z_{\rm inf,i}-f_{\rm out}*\dot{M}_{\rm ej,i}.
\end{equation}
Here $Z_{\rm i}$ denotes the abundance of element $i$ in the ISM while $Z_{\rm inf,i}$ denotes the abundance in the gas inflow. 
We assume pristine gas inflow and set $Z_{\rm in,i}$ to be 0. 
$\dot{M}_{\rm ej,i}$ represents the mass ejection rate of element $i$ from stars and $f_{\rm out}$ represents the mass fraction of stellar ejecta that will be driven out of the galaxy by galactic outflows. 

In this work we consider metal production by AGB winds, Type II and Type Ia supernova (SN-II and SN-Ia). 
Following the prescriptions given by \citet{tinsley1980}, the mass ejection rate of element $i$ by these processes is 
\begin{equation} \label{eq3}
\begin{aligned}
\dot{M}_{\rm ej,i}(t) & = \int_{0.85M_{\odot}}^{7M_{\odot}}M_{\rm i}^{\rm AGB}(M,Z_0)\psi(t-\tau_{\rm M})\phi(M){\rm dM}   \\
            & +  A'k\int_{\tau_{8M_{\odot}}}^{\tau_{0.85M_{\odot}}}M_{\rm i}^{Ia}\psi(t-\tau){\rm DTD}(\tau){\rm d\tau} \\
            & +  (1-A)\int_{7M_{\odot}}^{16M_{\odot}}M_{\rm i}^{II}(M,Z_0)\psi(t-\tau_{\rm M})\phi(M){\rm dM}     \\
            & +  \int_{16M_{\odot}}^{M_{\rm max}}M_{\rm i}^{II}(M,Z_0)\psi(t-\tau_{\rm M})\phi(M){\rm dM}.
\end{aligned}
\end{equation}
The first term accounts for the mass ejected by AGB stars while the second term denotes the contribution of SN-Ia with an analytic 
Delay Time Distribution (DTD) adopted from \citet{maoz2012}. The coefficient $A$ represents the percentage of stars in the mass range $3-16\ {\rm M_{\odot}}$ that end their lives with SN-Ia. We adopt an value of $0.028$ for $A$ which is similar to that commonly used in the literature \citep{greggio2005,arrigoni2010}. The other coefficient $A'$ is slightly different from $A$, representing the percentage of stars at full mass range that are SN-Ia progenitors. These two coefficients are related by $A'=A\times f_{3-16}$ \citep{arrigoni2010}, where $f_{3-16}$ represents the number fraction of stars that have masses between 3 and $16\ {\rm M_{\odot}}$. The coefficient $k$ in the second term represents the number of stars of a $1\ M_{\odot}$ single stellar population (SSP). The third and fourth term 
describe the mass ejected by SN-II after subtracting those stars ending in SN-Ia. The mass loss from these stellar objects is assumed to be ejected at the end of the lifetimes of the stars. 
The lifetime table for different masses of stars based on stellar tracks of the Padua library is taken from \citet{portinari1998}. For the initial mass function (IMF), we adopt a bimodal function with different slopes at low and high stellar masses:
\[
\phi(M) = 
\begin{cases}
a{\rm M}_*^{-\alpha1} &\text{if } {\rm M_*}<0.5{\rm M_{\odot}},\\
b{\rm M}_*^{-\alpha2} &\text{if } {\rm M_*}>0.5{\rm M_{\odot}}.
\end{cases}
\]
The slopes at the low ($\alpha1$) and the high ($\alpha2$) mass ends are set to be free parameters in the model. The fiducial values of $\alpha1$ and $\alpha2$ are 1.3 and 2.3 according to the Kroupa-IMF \citep{kroupa2001}, respectively. An IMF with $\alpha1>1.3$ is usually called bottom-heavy, while an IMF with $\alpha2>2.3$ is called a top-light. 

It should be noted that the mass ejection in equation \ref{eq2} includes the unprocessed metals as well as the newly {synthesized} metals. 
The former component is the metals that were locked into the progenitor star when it formed and therefore are related to the metallicity at the moment of formation.
The latter component indicates the metals {synthesized} during the lifetime of the star commonly known as the effective yield. For AGB winds, we take the metallicity-dependent effective yield table calculated by \citet{ventura2013} and
calculate unprocessed metals according to the metallicity when the progenitor of an AGB star is formed. 
We also test the model with alternative AGB yields by \citet{marigo2001}. We could verify that the difference in AGB yields does not affect the model results significantly, since SN-II ejecta dominate the metal enrichment at most times.  
For SN-II, we adopt the metallicity-dependent total mass ejection proposed by \citet{portinari1998} who {utilized} the SN nucleosynthesis model by \citet{woosley1995}. 
The unprocessed metals are already included using the metallicity of the grid.
The yields of individual elements by SN depend strongly on nucleosynthesis models \citep{thomas1998,thomas1999}, but this dependence is less relevant for the total metallicity discussed here.  
For SN-Ia, we adopt the widely-used spherically symmetric `W7' models with the yield tabulated by \citet{iwamoto1999}. Unlike the AGB winds and SN-II, the yield of SN-Ia is independent of the initial mass and metallicity of the progenitor star. The mass ejected by SN-Ia is a constant of $1.3\ {\rm M}_{\odot}$. Due to the uncertainties of the nature of SN-Ia progenitors \citep{greggio2005}, many empirical DTDs are proposed to match the observed SN-Ia rates \citep{strolger2004,matteucci2006,maoz2012}. 
In this work, we adopt the power-law DTD proposed by \citet{maoz2012}.   
To account for a possible time delay in the mixing of stellar ejecta with the ISM, we introduce a further parameter $t_{\rm delay}$.

To {summarize}, we have eight free parameters in our chemical evolution model:
\begin{itemize}
\item The coefficient and exponent of the KS law, $A_{\rm ks}$ and $n_{\rm ks}$.
\item The initial inflow strength, $A_{\rm inf}$ and its declining time scale $\tau_{\rm inf}$.
\item The outflow fraction $f_{\rm out}$
\item The IMF slope at the low mass and high mass ends, $\alpha1$ and $\alpha2$.
\item The time delay for the metal mixing $t_{\rm delay}$.
\end{itemize}
We assume that only stellar ejecta leave the galaxy in the outflow without interaction with the ISM of the galaxy. 
Since the metallicity of the stellar ejecta is usually much higher than the metallicity of the ISM, allowing other gas in the ISM leave the galaxy will not significantly enhance the effectiveness of suppressing the metal enrichment by the outflow. 
Considering the kinematic energy in the AGB winds, it may not be adequate to allow AGB winds out of the galaxy. We therefore also explore models with AGB winds excluded from the outflow presented in the discussion section. In the following subsection, we will explore the parameter space of these parameters to understand the role they play in the chemical evolution model. In addition, we also explore a possible time dependence of these parameters in our model. As we will show later, this time dependence turns out to be necessary to explain the dramatic metallicity evolution observed in low mass star-forming galaxies. 

\subsection{Exploring the parameter space}
As we mentioned above, there are eight free parameters in our chemical evolution model. Here we provide a brief analysis of how these parameters
affect the metal enrichment and the potential degeneracies between them. 
The ultimate aim is to identify which parameter combination is likely to reproduce the discrepancy in gas and stellar metallicity observed in star forming galaxies (\textsection2.3).
For simplicity, we only consider a single inflow process as scenario for the mass assembly of star-forming galaxies. As will be shown later, a single inflow process is adequate to explain most of the observations, such as the {MZR} and the {mass-SFR} (main sequence) relation. Multiple inflow episodes may be needed to reproduce more detailed structure of galaxies, such as the thick and thin disks of the Milk Way \citep{chiappini1997}. 
The time span of the model is set to be the age of the universe, 13.7 Gyr, with a grid step size of 0.1 Gyr.
To compare with the observed gas and mass-weighted stellar metallicities, we adopt the final oxygen abundance as the model gas metallicity and the mass-weighted metallicity of the composite stellar population at different ages as the 
model stellar metallicity. The stellar mass of an SSP component of the composite populations at any epoch is the current mass in stars at that epoch, accounting for the loss of massive and intermediate-mass stars due to stellar evolution.

A fiducial model is adopted with the parameter setting listed in Table 1. 
When exploring each parameter, the other parameters are set to be identical to the fiducial model. 
The fiducial model is a typical accretion box model with KS law parameters set to be the original value, IMF set to be bimodal Kroupa IMF, and no outflow. The final stellar mass of the fiducial model is $10^{10.25}{\rm M_{\odot}}$. The parameters are assumed to be constant with time. We will also explore models with time dependent parameters.
In the time-dependent model, we fix the value of the parameter at redshift 0 to be identical to the fiducial model while 
setting the initial value at the beginning as free. 

\begin{table*}
\caption{Parameter value of the fiducial chemical evolution model.}
\label{table1}
\centering
\begin{tabular}{l c c c c c c c c}
\hline\hline
$A_{\rm ks}$ & $n_{\rm ks}$ & $A_{\rm inf}$ & $\tau_{\rm inf}$ & $f_{\rm out}$ & $t_{\rm out}$ & $\alpha1$ & $\alpha2$ & $t_{\rm delay}$ \\
 & & ${\rm M_{\odot} yr}^{-1}$ & Gyr & & Gyr & & & Gyr \\
\hline
1 & 1.5 & 3.16 & 10 & 0 & 0 & 1.3 & 2.3 & 0 \\
\hline
\end{tabular}\\  
\end{table*}

\subsubsection{Star formation description}
The first two parameters we discuss are the coefficient ($A_{\rm ks}$) and the exponent ($n_{\rm ks}$) of the KS law. 
An important quantity determined by the KS law is the SFE, i.e. the SFR per mass unit of gas.
The SFE is positively correlated to $A_{\rm ks}$ and $a_{\rm ks}$. 
Since the metal production is directly related to the star formation activity, 
the metallicity of a galaxy, 
i.e.\ the mass or number ratio between chemical elements and gas, should be sensitive to the variation of the SFE. 
Therefore, the two parameters of the KS law will affect the model metallicity through changes of the SFE.
 
Figure 5 shows the time evolution of the gas metallicity, the stellar metallicity, and the SFR for various values of $A_{\rm ks}$ shown as solid lines. 
The red solid line represents the predictions of the fiducial model with the parameters listed in Table 1. 
It can be seen that the difference between today's gas and stellar metallicities in the fiducial model is $<0.1$~dex, hence much smaller than what is found in observations. 
In the models with high $A_{\rm ks}$, both the gas and the stellar metallicities increase rapidly at early times and then begin to saturate and barely change at later times. The final gas and stellar metallicities are relatively high with no significant difference between them.
In contrast, in the model with low $A_{\rm ks}$, the enrichment of metals is suppressed with low gas and stellar metallicities.
Since with lower $A_{\rm ks}$ fewer stars are formed at early times, the final stellar metallicity is still close to the gas metallicity with a difference less than 0.4 dex. {This difference may be large enough to match the lower limit of the observed metallicity difference when adopting the gas metallicity derived through the N2 method. However, such a low $A_{\rm ks}$ (1\% of the original value) would imply a star-formation law that deviates substantially from the KS law, which seems contrived. 
Moreover, in the scenario of varying $A_{\rm ks}$, the change of gas metallicity is similar to the stellar metallicity and it is difficult to
reproduce the distinctive dynamic range of gas metallicity (0.3~dex) and stellar metallicity (0.8~dex) simultaneously. }
It is worth noting that the final gas metallicity does not change significantly when $A_{\rm ks}$ increases from 0.1 to 1. This suggests that the gas metallicity becomes insensitive to the variation of the SFE at high SFE. 

In the time-dependent scenario we assume that $A_{\rm ks}$ increases linearly with cosmic time.
The dashed lines in Figure 5 show the metal enrichment and star formation histories predicted by such a time-dependent chemical evolution model. The initial value of $A_{\rm ks}$, $A_{\rm ks,i}$, is shown in the legend. 
At early times, the gas and stellar metallicities are suppressed owing to the low value of $A_{\rm ks}$. However, at late times, both the gas metallicity and the stellar metallicity increase rapidly with the increasing $A_{\rm ks}$ and reach a similar value to the fiducial model. Although a dramatic evolution of the SFE (i.e.\ a lower SFE at early times) could lead to a dramatic evolution of metallicity (i.e.\ a lower metallicity at early times), the number of stars formed at early times in low metallicity environments will also be dramatically reduced. This cancels the lower metallicity effect and hence leaves the mass-weighted stellar metallicity virtually unchanged. 

Figure 6 shows the evolution of the metallicity and the SFR predicted by models with varying $n_{\rm ks}$ (solid) and varying $n_{\rm ks,i}$ (dashed). 
It can be seen that both the metal enrichment history and the star formation history change with varying $n_{\rm ks}$ and $n_{\rm ks,i}$ in a similar way as with varying $A_{\rm ks}$ and $A_{\rm ks,i}$ (Figure 5). This is due to the fact that $A_{\rm ks}$ and $n_{\rm ks}$ regulate the metal enrichment processes through the same parameter, SFE. 
{Similar to the models with varying $A_{\rm ks}$ in Figure 5, the models with varying $n_{\rm ks}$ may be able to reproduce the metallicity difference when adopting the gas metallicity derived through the N2 method, but they cannot easily reproduce the dynamic range of gas and stellar metallicities simultaneously.}  

\begin{figure*}
\centering
\includegraphics[width=18cm]{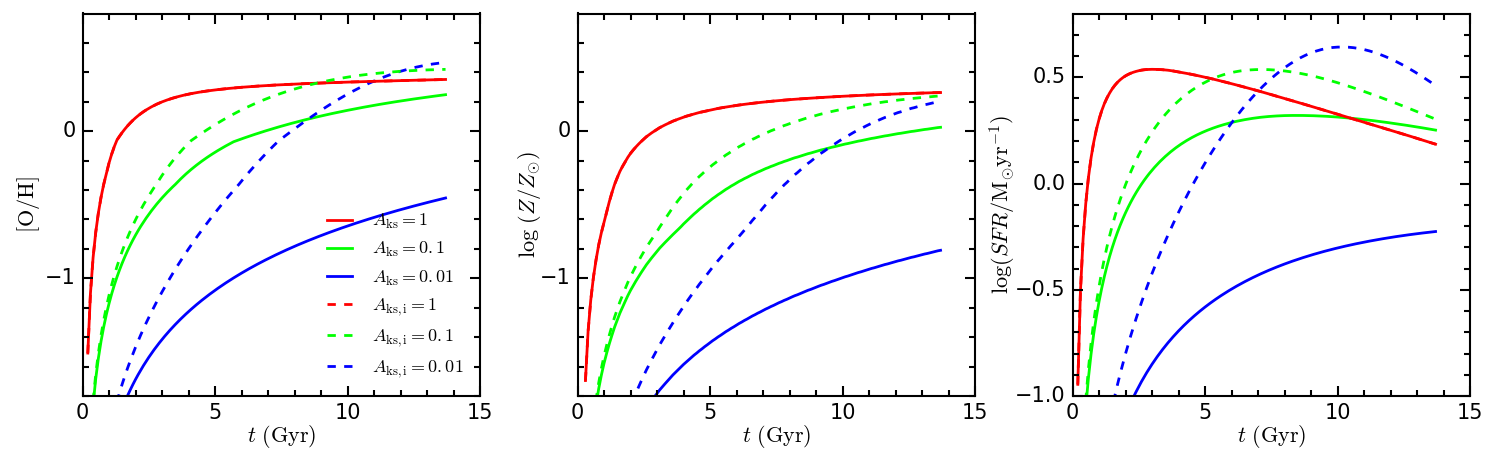}
\caption{Evolution of gas metallicity, stellar metallicity, and SFR as predicted by our chemical evolution model. Solid lines are models with a star formation law coefficient $A_{\rm ks}$ constant with time, while dashed lines are models with time-dependent $A_{\rm ks}$ in which we vary the initial star formation law coefficient $A_{\rm ks,i}$. Different colours represent the models with varying $A_{\rm ks}$ and $A_{\rm ks,i}$ as shown in the legend. The values of other parameters are fixed to the fiducial model of Table 1. The red dashed line can not be seen due to overlap with the solid red line.  
}
\label{figure5}
\end{figure*}

\begin{figure*}
\centering
\includegraphics[width=18cm]{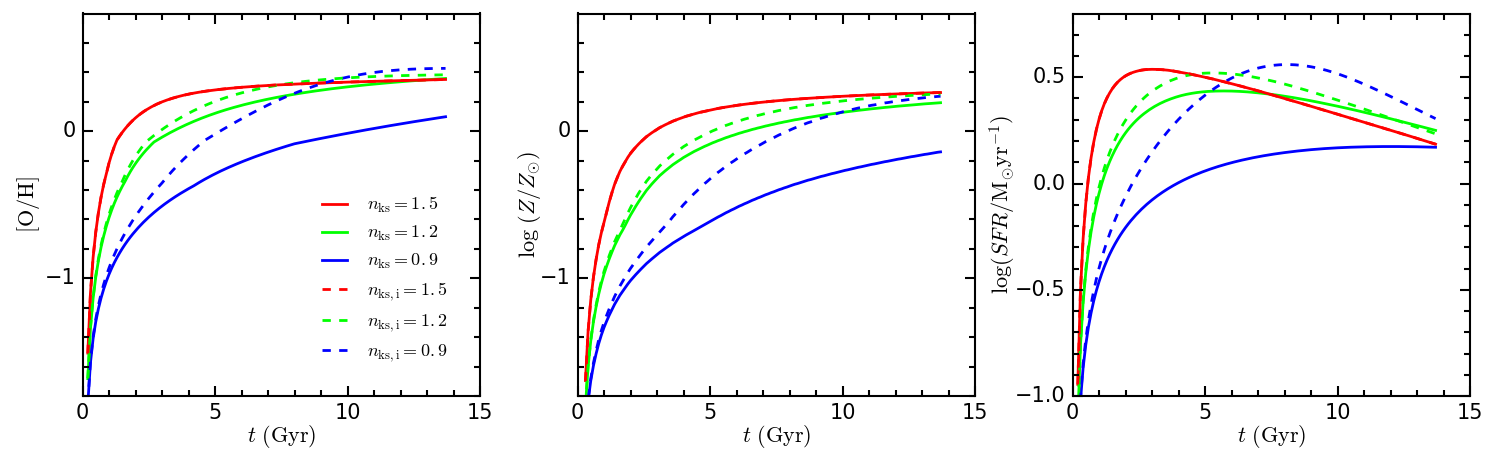}
\caption{Similar to Figure 5 except that the varying parameter is the star formation law exponent, $n_{\rm ks}$ (solid, constant with time) and $n_{\rm ks,i}$ (dashed, time dependent). 
}
\label{figure6}
\end{figure*}

\subsubsection{Gas inflow}
The evolution of gas metallicity, stellar metallicity, and SFR as predicted by chemical evolution models with variable initial inflow rate $A_{\rm inf}$ and inflow time-scale $\tau_{\rm inf}$ are shown in Figure 7.
It turns out that the gas and stellar metallicities increase with a higher inflow rate, especially at early times. However, the final gas metallicity hardly changes with $A_{\rm inf}$ increasing from 0.1 to 1 ${\rm M_{\odot}yr^{-1}}$. This confirms our interpretation that the variation of gas metallicity caused by a variation in SFE is negligible when SFE is high. 
These features are similar to those found in Figure 5 with varying $A_{\rm ks}$, and Figure 6 with varying $n_{\rm ks}$. We note that the SFE in our chemical evolution model is actually regulated by three parameters, the star formation law coefficient $A_{\rm ks}$, the exponent $n_{\rm ks}$, and the inflow rate $A_{\rm inf}$. This is the reason why we find similar trends of the metal enrichment history and the star formation history with changes of these parameters. 
A high degree of degeneracy between these parameters ought to be expected in models reproducing the observed gas metallicity.

In terms of inflow time scale, although the star formation history changes significantly with varying $\tau_{\rm inf}$, the metal enrichment history does not change significantly. 
With an inflow time scale of 1~Gyr, the model predicts a gas metallicity that is higher by $\sim0.3$~dex compared to the fiducial model, and a stellar metallicity that barely changes. Nevertheless, the largest difference between the late time gas and stellar metallicities achieved by varying $\tau_{\inf}$ is $\sim0.4$ dex, {which barely matches the lower limit of the observed value. Moreover, the much higher gas metallicity when adopting a low $\tau_{\inf}$ would require other extreme scenarios to match the observation.}

\begin{figure*}
\centering
\includegraphics[width=18cm]{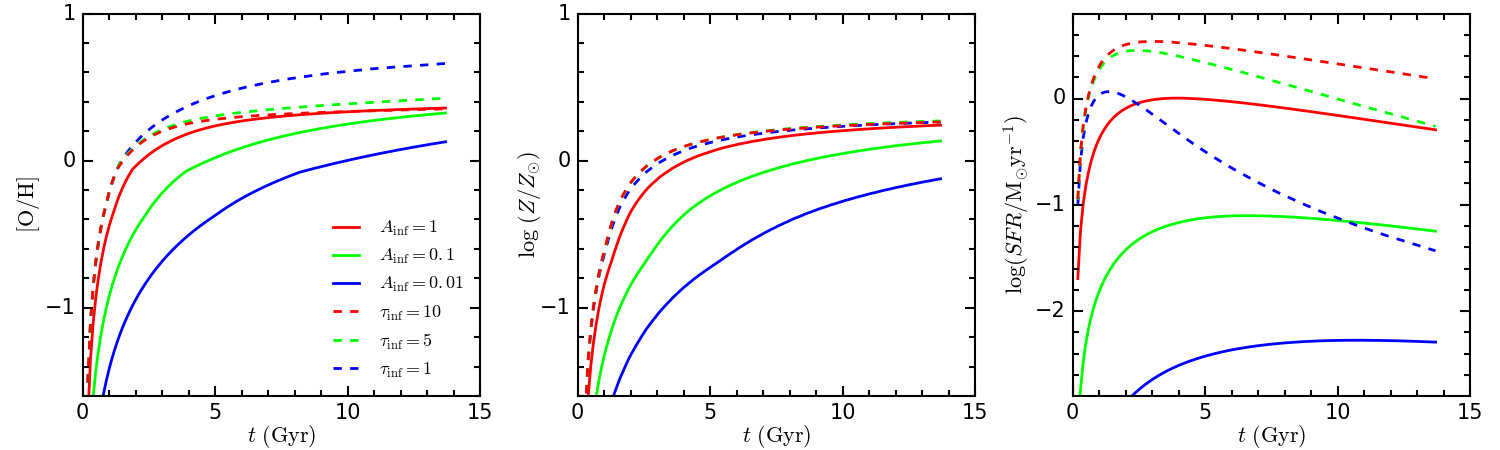}
\caption{Similar to Figure 5 but varying initial inflow strength $A_{\rm inf}$ (solid) and time-scale (dashed).
}
\label{Figure7}
\end{figure*}

\subsubsection{Gas outflow}
Figure 8 shows the variation in metal enrichment history and star formation history induced by changing the outflow fraction $f_{\rm out}$ (solid lines). Here we assume that the outflow contains all stellar ejecta, including the material ejected by AGB winds. We will discuss the effect of excluding the AGB winds from the outflow in the discussion section. 
As expected, models with a higher outflow fraction give lower gas and stellar metallicities. However, the difference between gas and stellar metallicity remains below 0.3 dex. The star formation history changes slightly for the different outflow fractions. With higher $f_{\rm out}$, the SFR tends to be slightly lower due to the lower gas mass resulting from the fact that less stellar ejecta are recycled back to the ISM. 

Since the star formation history changes slightly with different $f_{\rm out}$, a dramatic evolution of the outflow fraction with strong outflow at early times could possibly reproduce the lower stellar metallicity relative to the gas metallicity. The dashed lines in Figure 8 represent the time-dependent outflow model. We use a modified linear function to describe the time evolution of the outflow fraction as illustrated in Figure 9. 
Before a transition time point $t_{\rm out}$, the initial outflow fraction $f_{\rm out,i}$ is set to be 100 per cent. The gas and stellar metallicity are zero during this period. After the transition at time point $t_{\rm out}$, the outflow fraction begins to {decrease} linearly to the present value $f_{\rm out,p}$. 
It can be seen that the gas metallicity at late epochs increases more rapidly than the stellar metallicity.
As a result, the difference between the gas and stellar metallicity reaches 0.6~dex when 
$t_{\rm out}=6$ Gyr, i.e. the outflow fraction is 
100 per cent for the first 6 Gyr. Since the outflow fraction is assumed to be 100 per cent before $t_{\rm out}$, the metal enrichment process is entirely suppressed before this time point. With higher $t_{\rm out}$, the metal enrichment process is suppressed for a longer time. 
Thus, more stars form in an extremely low metallicity environment at early epochs, the lower will be the mass-weighted stellar metallicity at late times. The gas metallicity, instead, could easily catch up after $t_{\rm out}$, and thus is not highly dependent on this parameter. The time-dependent outflow model with a strong outflow at early times is therefore the first scenario that could possibly reproduce the observed gas and stellar metallicity at the same time. 

\begin{figure*}
\centering
\includegraphics[width=18cm]{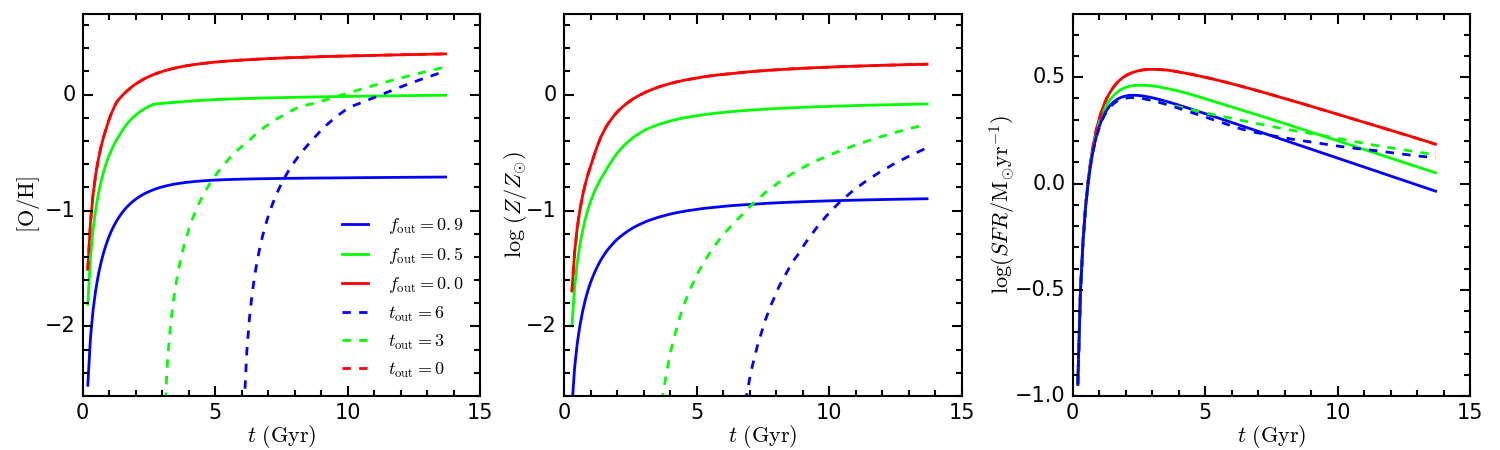}
\caption{Similar to Figure 5 but varying the {metal} outflow fraction $f_{\rm out}$ (solid) and the transition time $t_{\rm out}$ (dashed).}
\label{Figure8}
\end{figure*}

\begin{figure}
\centering
\includegraphics[width=\columnwidth,viewport=25 10 660 620,clip]{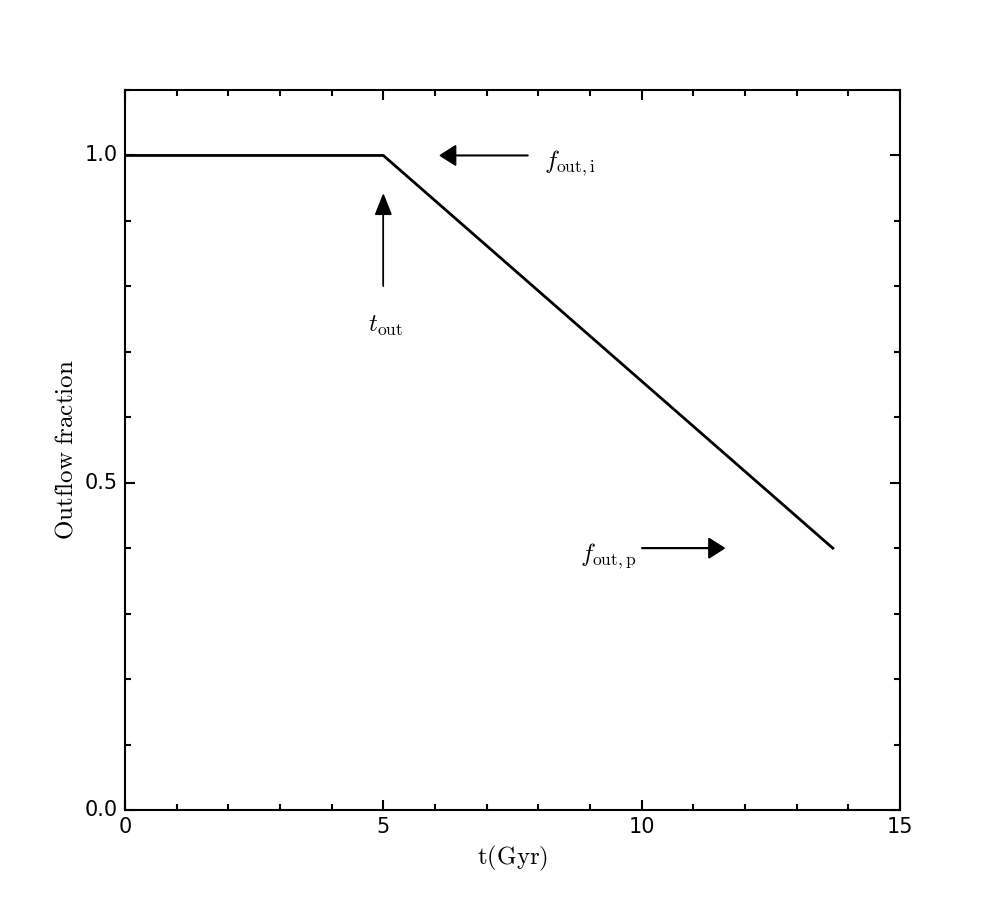}
\caption{Schematic plot to illustrate the function adopted for the time evolution of the {metal} outflow fraction. The {initial metal outflow fraction $f_{\rm out,i}$}, present {metal} outflow fraction $f_{\rm out,p}$, and the transition time $t_{\rm out}$ are marked in the plot. 
}
\label{Figure9}
\end{figure}

\subsubsection{IMF}
We further explore a chemical evolution model with varying IMF slope at the low mass end ($\alpha1$) and at the high mass end ($\alpha2$) as shown in Figure 10 and Figure 11, respectively. 
Similar to the other parameters, solid lines 
indicate the time-constant model while dashed lines are for models assuming a time-dependent IMF slope. 
It can be seen that the model with steeper IMF slope both at the low mass end (bottom heavy IMF) and at the high mass end (top light IMF)
predict lower gas and stellar metallicities. This can be understood as massive stars dominate the metal production. A steeper IMF slope implies a lower number of massive stars, therefore a lower efficiency in metal production and hence lower metallicity. 
However, because both gas and stellar metallicity are very sensitive to the IMF slope, changing the IMF slope alone does not produce a large enough difference between gas and stellar metallicity. 

Since steepening the IMF slope lowers the metallicity efficiently while not changing the star formation history significantly, we can expect a model with a steep IMF at early times but a normal IMF at late times to produce 
a much lower metallicity in the stars than that in the gas. The dashed lines in Figure 10 and Figure 11 show such a time-dependent IMF model. 
In this model, the IMF slope changes linearly with time and becomes a bimodal Kroupa IMF at the present epoch.
The initial IMF slope $\alpha1_{\rm i}$ and $\alpha2_{\rm i}$ are set to be free parameters. 
As expected, the models with a steeper initial IMF slope either at the low mass or the high mass end predict a stellar metallicity significantly lower than the gas metallicity. Despite dramatic variations in stellar metallicity, the gas metallicity tends again to catch up at late epochs. Therefore, we consider a time-dependent variable IMF with a steep slope at early times as a further 
effective scenario to explain the observed metallicity properties of local star-forming galaxies. 
 
\begin{figure*}
\centering
\includegraphics[width=18cm]{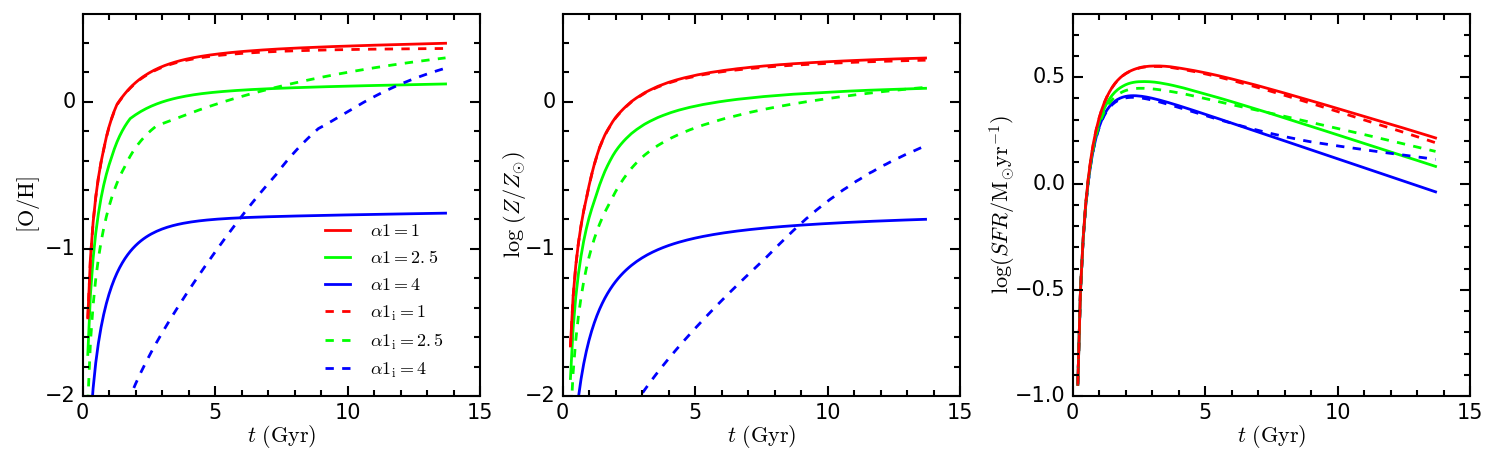}
\caption{Similar to Figure 5 but varying the IMF slope at the low mass end $\alpha1$. Solid line shows the model with time-constant $\alpha1$ while dashed lines are for models with time-dependent $\alpha1$ in which we change the initial IMF slope $\alpha1_{\rm i}$.
}
\label{Figure10}
\end{figure*} 

\begin{figure*}
\centering
\includegraphics[width=18cm]{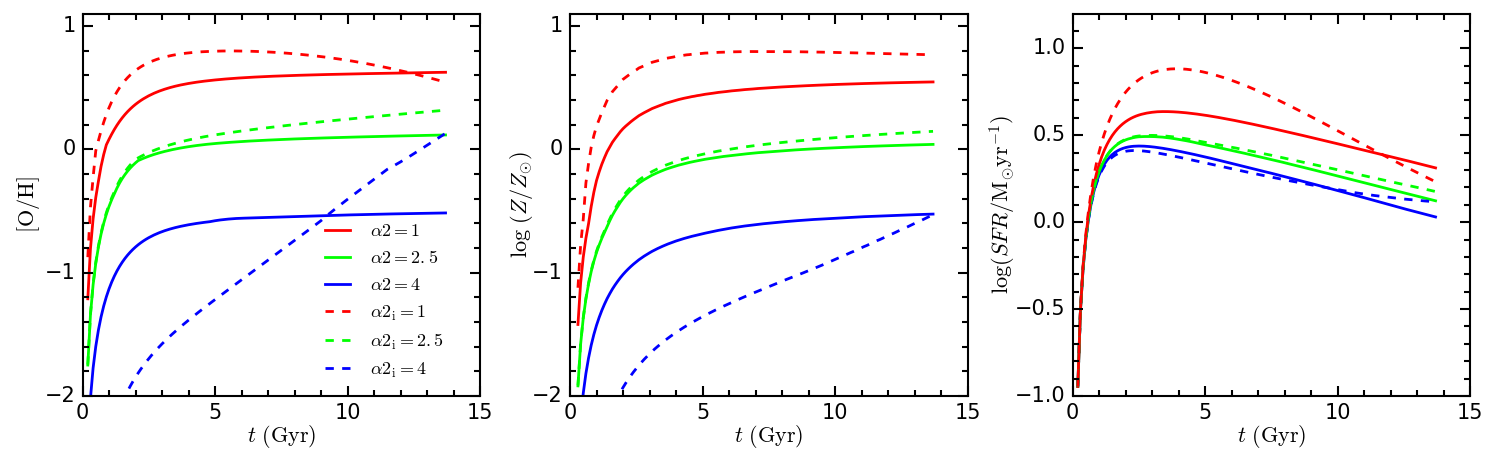}
\caption{Similar to Figure 5 but varying the IMF slope at high mass end $\alpha2$. Solid line shows the model with time-constant $\alpha2$ while dashed lines are for models with time-dependent $\alpha2$ in which we change the initial IMF slope $\alpha2_{\rm i}$.
}
\label{figure11}
\end{figure*} 

\subsubsection{Mixing delay}
In the previous models we assume instantaneous mixing of the metals in the ISM. In this section we explore chemical evolution models with a delayed mixing of the stellar ejecta instead. Figure 12 shows chemical evolution models with various delay-times $t_{\rm delay}$ for the mixing. {It is interesting to note that the gas metallicity barely changes while the stellar metallicity tends to be slightly lower in models with longer (initial) time-delay. With a time-delay of 4~Gyr, the difference in gas and stellar metallicity reaches $\sim 0.4$ dex which is close to the lower limit of the observed value.} 
By exploring the chemical evolution of the Milky Way in [Mg/Fe] versus [Fe/H], however, \citet{thomas1998} found that a model with a time-delay of the order of 10$^8$ yr matches the observational data best. As we can see from Figure 12, this order of time-delay only introduces a difference between gas and stellar metallicity similar to the fiducial model.   
{Therefore, we conclude that a time-delay in metal mixing below 1~Gyr cannot be the origin of the observed metallicity difference between stars and gas.}   

\begin{figure*}
\centering
\includegraphics[width=18cm]{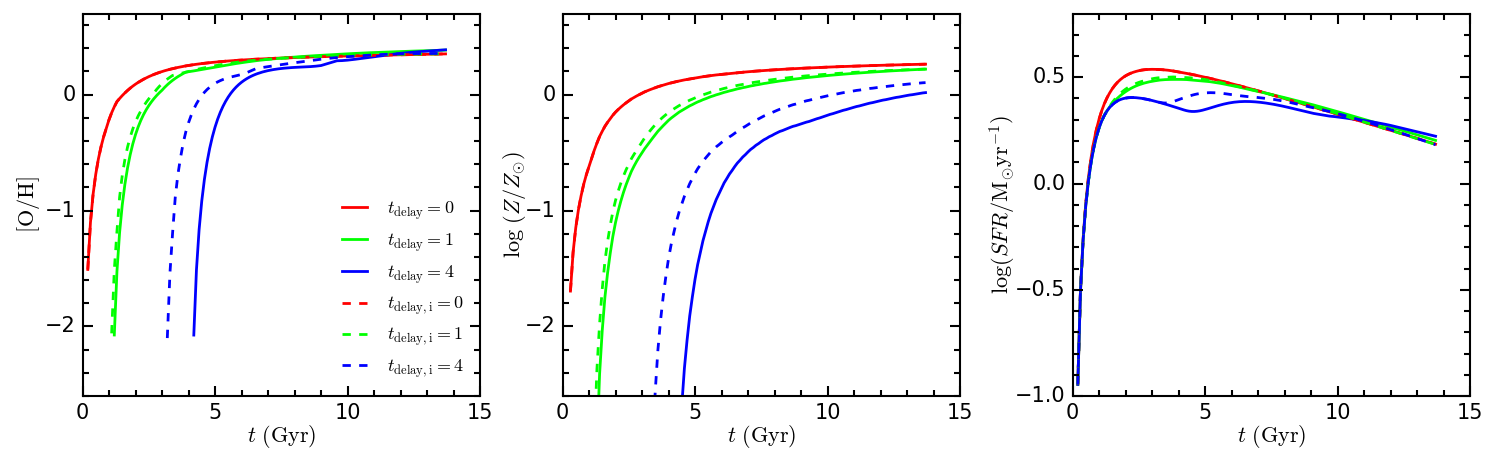}
\caption{Similar to Figure 5 but varying the time-delay $t_{\rm delay}$ (solid) and $t_{\rm delay,i}$ (dashed) in the mixing the stellar ejecta with the ISM. 
}
\label{Figure12}
\end{figure*} 

\subsection{Summary of the modelling approach}

After exploring the variation of several parameters in our chemical evolution model, we now proceed to discuss possible degeneracies between them. The KS law coefficient $A_{\rm ks}$, the power law index $n_{\rm ks}$ and the initial inflow strength $A_{\rm inf}$
all regulate the metal production by changing the SFE and therefore affect the gas 
and stellar metallicities in similar ways. We find that the variation of these parameters does not generate the discrepancy between gas and stellar metallicity observed in local {star-forming} galaxies.
Since the observed dynamic range in the gas metallicity measurements is much narrower than the range in stellar metallicity, these three parameters could
only contribute to the residual gas metallicity variation. %
The other inflow parameter, inflow time-scale, is mainly constrained by the {mass-SFR} relation and therefore does not introduce further degeneracies.

The other three parameters explored are the outflow fraction $f_{\rm out}$ and the IMF slope at the low mass $\alpha1$ and the high mass end 
$\alpha2$. All three directly regulate the metal production
without affecting the star formation history. These parameters, together with late-time outflow fraction and IMF slope, are degenerate in gas metallicity. Therefore, the {MZR$_{\rm gas}$} can be generated by a combination of any of these parameters. The time-delay $t_{\rm delay}$ in the mixing of the stellar ejecta with the ISM, instead, does not affect the late-time gas metallicity and does not introduce any significant variation in the stellar metallicity.   
As we discussed above, a 
time-dependent outflow fraction or IMF slope, instead, are both successful in reproducing the observed difference in gas and stellar metallicity. Therefore, the outflow transition time $t_{\rm out}$ and the early-time IMF slopes $\alpha1_{\rm i}$, $\alpha2_{\rm i}$) are the major parameters under consideration to reproduce the observations. 

\section{Results}
In the previous sections we explored the model parameter space and discussed the potential degeneracies. 
In spite of the large range of model parameters, it turns out to be surprisingly difficult to reproduce the observed large difference between gas and stellar metallicities. Hence, the final model is relatively well-constrained.
Among all the parameters, we only find two scenarios that are able to explain the observations, i.e.\
either a strong metal outflow or a steep IMF slope at early times in the chemical enrichment history. 
In this part we first illustrate how the normal chemical evolution models with parameters invariable with time fail to reproduce the observations and then introduce the two successful models that invoke a time dependence on the model parameters. 
We fix the KS law power index to be 1.5 which is close to the value of 1.4 found in \citet{kennicutt1998}.

Figure 13 shows the predicted {MZR} for a series of fiducial models with varying inflow strength $A_{\rm inf}$. 
The dynamic range of galaxy stellar mass is obtained through adopting a range of different initial inflow strengths $A_{\rm inf}$. More massive galaxies have a higher $A_{\rm inf}$. 
It can be seen that the normal accreting box model predicts higher metallicities both in the gas and in stars, as well as much flatter {MZRs} compared to the observations. The chemical evolution model with normal parameter settings (i.e.\ KS star-formation law, Kroupa IMF, and no outflow) is too efficient in metal production and therefore yields metallicities that are too high.  

As suggested in the literature, there are three possible scenarios that are capable of explaining the {MZR$_{\rm gas}$}. One scenario uses a mass-dependent SFE, assuming that more massive galaxies have a higher SFE and therefore higher gas metallicities \citep{brooks2007,calura2009}. Alternatively or additionally, stronger outflows in less massive galaxies due to the lower gravitational potential well can produce the {MZR$_{\rm gas}$}. Lastly, less massive galaxies would be less metal-rich, if their stellar populations are characterised by steeper IMF slopes.
Based on these three scenarios, we tune our chemical evolution model to match the observed gas metallicity-relation as derived through the R23 and N2 methods as shown by solid lines in Figure 14 and Figure 15, respectively. The parameters adopted for the models in Figure 14 and Figure 15 are shown in Table~2. The grey shaded region indicates the 16th and 84th percentiles, while the black shaded region indicates the error of the median metallicity in each mass bin. As expected, the models based on these three scenarios are all equally successful in reproducing the observed {MZR$_{\rm gas}$}. The exact values of the model parameters depend on the choice of gas metallicity calibration. 

Despite the good match in gas metallicity, however, these models are not able to reproduce the observed {MZR$_{\rm star}$}. 
When adopting the gas metallicity derived through the R23 method, the models that reproduce the observed {MZR$_{\rm gas}$} predict a stellar metallicity systematically higher than the observed value in the whole mass range (Figure~14). The {MZR$_{\rm star}$} predicted by these models also tends to be flatter compared to the observation. When the gas metallicity is measured through the N2 method, which leads to much lower metallicities than the R23 method, the models can in principle reproduce the observed value of stellar metallicity but still are not able to match the correct slope of the {MZR$_{\rm star}$} (Figure~15). 

In addition to this, the very low gas metallicities derived from the N2 method lead to parameter choices that are not well supported by observations. For the scenario of a mass-dependent SFE, the coefficient of the KS law is required to be as low as 1\% of the value calibrated by \citet{kennicutt1998}, in order to reproduce the (relatively low) N2 gas metallicities. Within the other scenarios either unrealistically strong outflows or an extremely steep IMF are needed to reach the relatively low gas metallicity observed. This seems inconsistent with the finding of only minor outflows in local star-forming galaxies \citep{concas2017} and the only mild IMF variations at the low mass end \citet{bastian2010}. On top of this, more massive early type galaxies tend to have a steeper IMF (if any IMF variation at all) \citep[e.g.,][]{conroy2012,labarbera2013}, which is opposite to the trend required here to reproduce the low N2 gas metallicities in low mass galaxies. To {summarize}, chemical evolution models with time-independent parameters are not able to reproduce the observed {MZR$_{\rm gas}$} and {MZR$_{\rm star}$} simultaneously.

\begin{figure*}
	\centering
	\includegraphics[width=16cm]{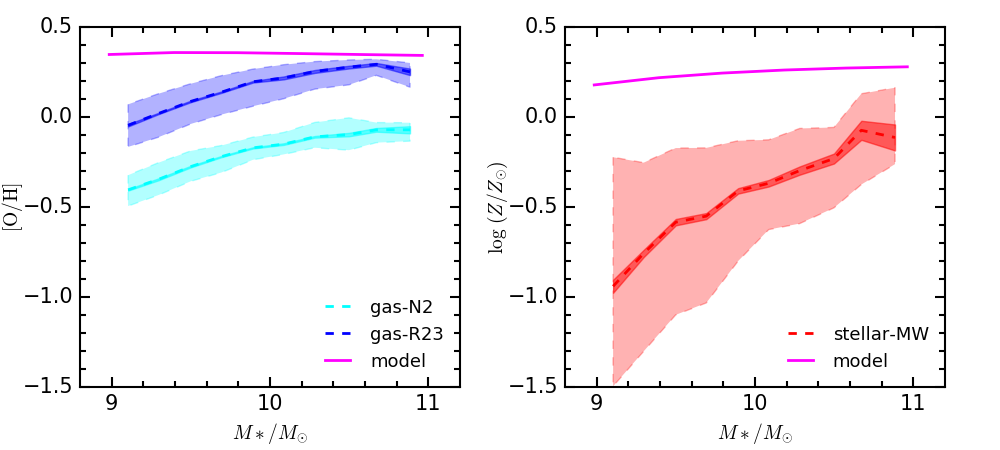}
	\caption{Comparison of the predictions from the fiducial accreting box chemical evolution model with observations in {MZR$_{\rm gas}$ and MZR$_{\rm star}$}.
	}
	\label{figure13}
\end{figure*}

\begin{figure*}
	\centering
	\includegraphics[width=16cm]{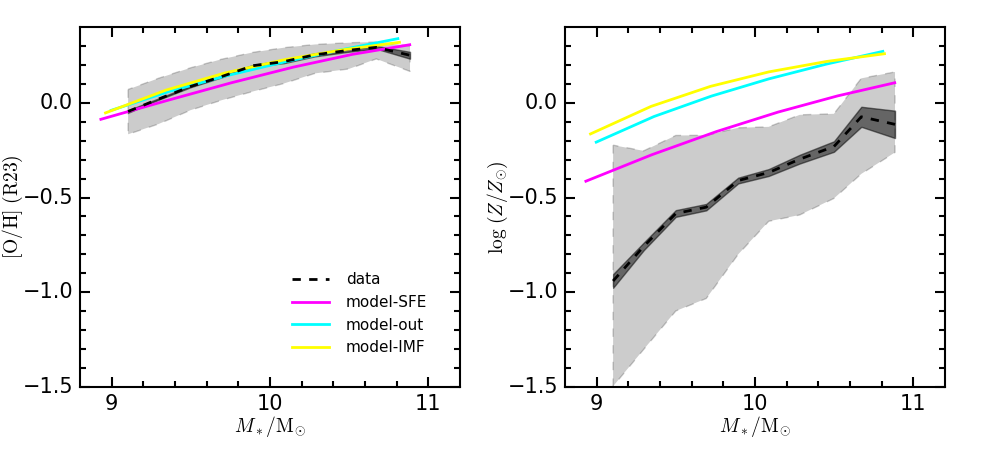}
	\caption{Comparison of the predictions from chemical evolution model that are tuned to match the observed {MZR$_{\rm gas}$}. There are three effective ways to reproduce the {MZR$_{\rm gas}$} by introducing a mass-dependent SFE, mass-dependent outflow strength, or mass-dependent IMF. The predictions of these three models are shown in different colours as indicated in the legend. The gas metallicity is derived from the R23 method.} 
	\label{figure14}
\end{figure*}

\begin{figure*}
	\centering
	\includegraphics[width=16cm]{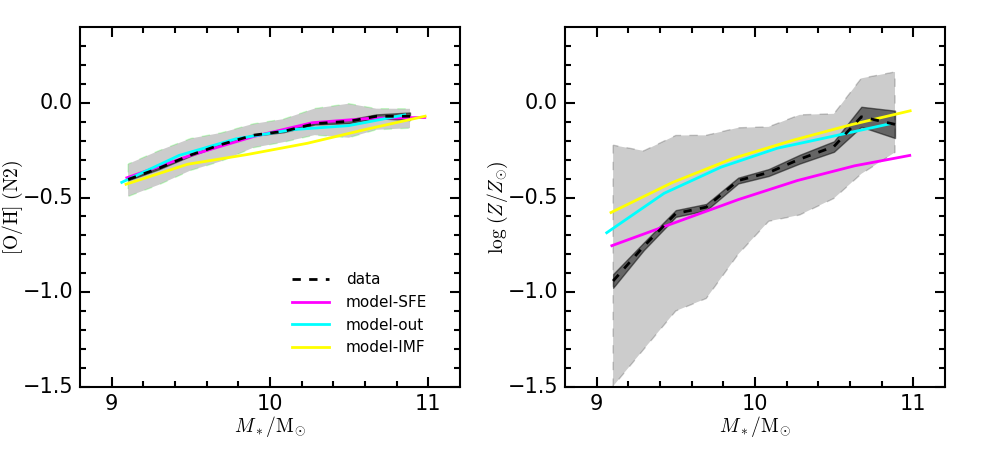}
	\caption{{Similar to Figure 14 but adopting gas metallicity derived by the empirical N2 method.}
	}
	\label{figure15}
\end{figure*}

\begin{table*}
	\caption{Parameter value of the normal chemical evolution models in Figure 13, Figure 14 and Figure 15. Parameter values in the square brackets are for the models of lowest and highest galaxy masses, respectively.}
	\label{table2}
	\centering
	\begin{tabular}{l c c c c c c c c c }
		\hline\hline
models &	$A_{\rm ks}$ & $n_{\rm ks}$ & $A_{\rm inf}$ & $\tau_{\rm inf}$ & $f_{\rm out}$ & $\alpha1$ & $\alpha2$ & $t_{\rm delay}$ \\
	&   & & ${\rm M_{\odot}yr^{-1}}$ & Gyr & & & & Gyr \\
\hline
fiducial & 1 & 1.5 & [0.20,15.85] & 10 & 0 & 1.3 & 2.3 & 0 \\
model-SFE-R23 & 0.06 & 1.5 &  [0.50,17.78] & 10 & 0 & 1.3  & 2.3 & 0 \\
model-out-R23 & 1 & 1.5 & [0.25,11.22] & 10 & [0.56,0.01] & 1.3 & 2.3 & 0 \\
model-IMF-R23 & 1 & 1.5 & [0.16,11.22] & 10 & 0 & [2.30,1.40] & 2.3 & 0 \\
model-SFE-N2 & 0.08 & 1.5 & [0.63,25.12] & 10 & 0.6 & 1.3  & 2.3 & 0 \\
model-out-N2 & 1 & 1.5 & [0.32,15.85] & 10 & [0.82,0.54] & 1.3 & 2.3 & 0 \\
model-IMF-N2 & 1 & 1.5 & [0.20,14.13] & 10 & 0 & [2.76,2.28] & 2.3 & 0 \\
\hline
	\end{tabular}\\  
\end{table*}

\subsection{Best-fitting models}
{In this section, we present the best-fitting chemical evolution models that reproduce both {MZRs} simultaneously. We add the {mass-SFR} relation as a further observational constraint to the model.} We seek best-fitting models that reproduce simultaneously today's gas metallicity, stellar metallicity, and star formation rate in star forming galaxies as a function of galaxy mass. For simplicity, here we focus on the gas metallicity derived through the R23 method. We will discuss the effect caused by the choice of gas metallicity calibration method in the discussion section.

\subsubsection{Variable metal outflow}
As we show in \textsection3.2, a model with a high metal outflow fraction at early times successfully produces the much lower metallicity of the stellar component. 
To match the observed {MZR and mass-SFR} relations, we generate a series of models with a range of initial inflow strengths
$A_{\rm inf}$ tuning the other parameters $A_{\rm ks}$, $\tau_{\rm inf}$, $f_{\rm out}$, and $t_{\rm out}$, as a linear function of $A_{\rm inf}$ (hence galaxy mass). 
Here {the metal outflow is time-dependent and} the AGB wind is assumed to be included in the outflow. 
Figure 16 shows the comparison between the {predictions} of these models with the observations in {MZR$_{\rm gas}$} (gas metallicities from the R23 method), the mass-weighted {MZR$_{\rm star}$} and the {mass-SFR} relation. 
The dashed line represents the observed median {MZRs}. The light and dark shaded regions indicate the 16th and 84th percentiles and the error of the median metallicities, respectively.
The evolution of the resulting metallicity, stellar mass, SFR, inflow rate and outflow fraction is illustrated in Figure 17. The adopted parameter ranges are shown in Table~3.

From Figure 16 it can be seen that the model with time-dependent outflow fraction successfully reproduces the observed {MZR$_{\rm gas}$, MZR$_{\rm star}$ and mass-SFR} relations. In this model, the steep {MZR$_{\rm star}$} is driven by a mass-dependent $t_{\rm out}$. Less-massive galaxies have higher $t_{\rm out}$ values, i.e.\ higher average metal outflow rates at early times and therefore lower average stellar metallicity. This solution seems plausible astro-physically, since the gravitation well is expected to be shallower at early times, particularly in low-mass systems. {In this scenario}, the relatively flat {MZR$_{\rm gas}$} is also mainly driven by the mass-dependent $t_{\rm out}$. The mass-dependent $A_{\rm inf}$, hence mass-dependent SFE, further contributes to the slope of {MZR$_{\rm gas}$}. It is worth noting that the flattening of {MZR$_{\rm gas}$} at the high-mass end is also well reproduced in our model. This is because the gas metallicity becomes saturated in massive galaxies due to the generally high SFE. To understand the underlying physics, including potential {environmental} effects, that drive the time-dependent outflow, a high resolution hydrodynamical simulation may be needed which is beyond the scope of this work. 
As to the origin of the {mass-SFR} relation, the mass-dependent inflow time-scale with higher $\tau_{\rm inf}$ in lower mass galaxies naturally produces the main sequence relation. This is consistent with the `downsizing' scenario in which less-massive galaxies are less efficient in converting gas into stars and therefore forming stars on a longer time scale compared to more massive objects \citep[e.g.,][]{cowie1996,thomas2005,noeske2007,thomas2010}.

\begin{figure*}
\centering
\includegraphics[width=17cm,viewport=0 10 700 640,clip]{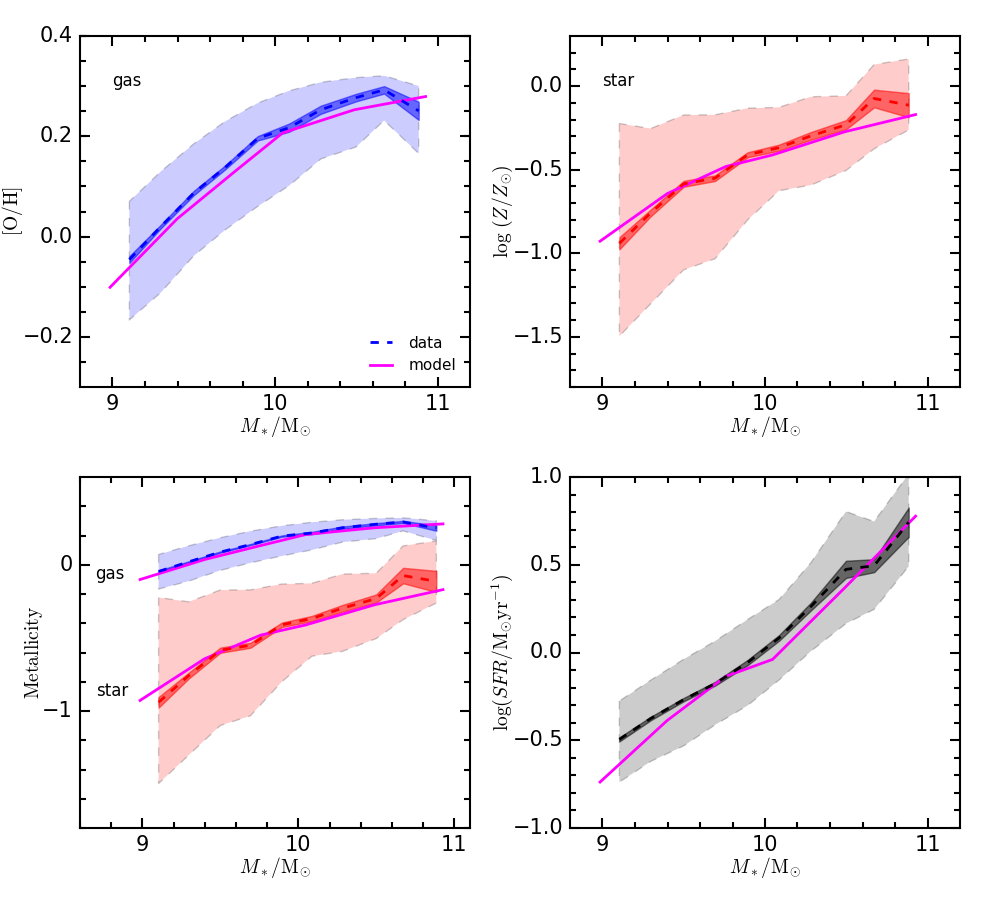}
\caption{Comparison between the finely-tuned variable outflow model with the observed {MZR$_{\rm gas}$, mass-weighted MZR$_{\rm star}$, and 
mass-SFR relation}. Dotted lines represent the median of the distribution, while the shaded regions denote the 16th to 84th percentiles.
{The dark shaded regions indicate the error of median metallicities at a given mass.}}

\label{figure16}
\end{figure*}

\begin{figure*}
\centering
\includegraphics[width=17cm,viewport=10 10 1100 640,clip]{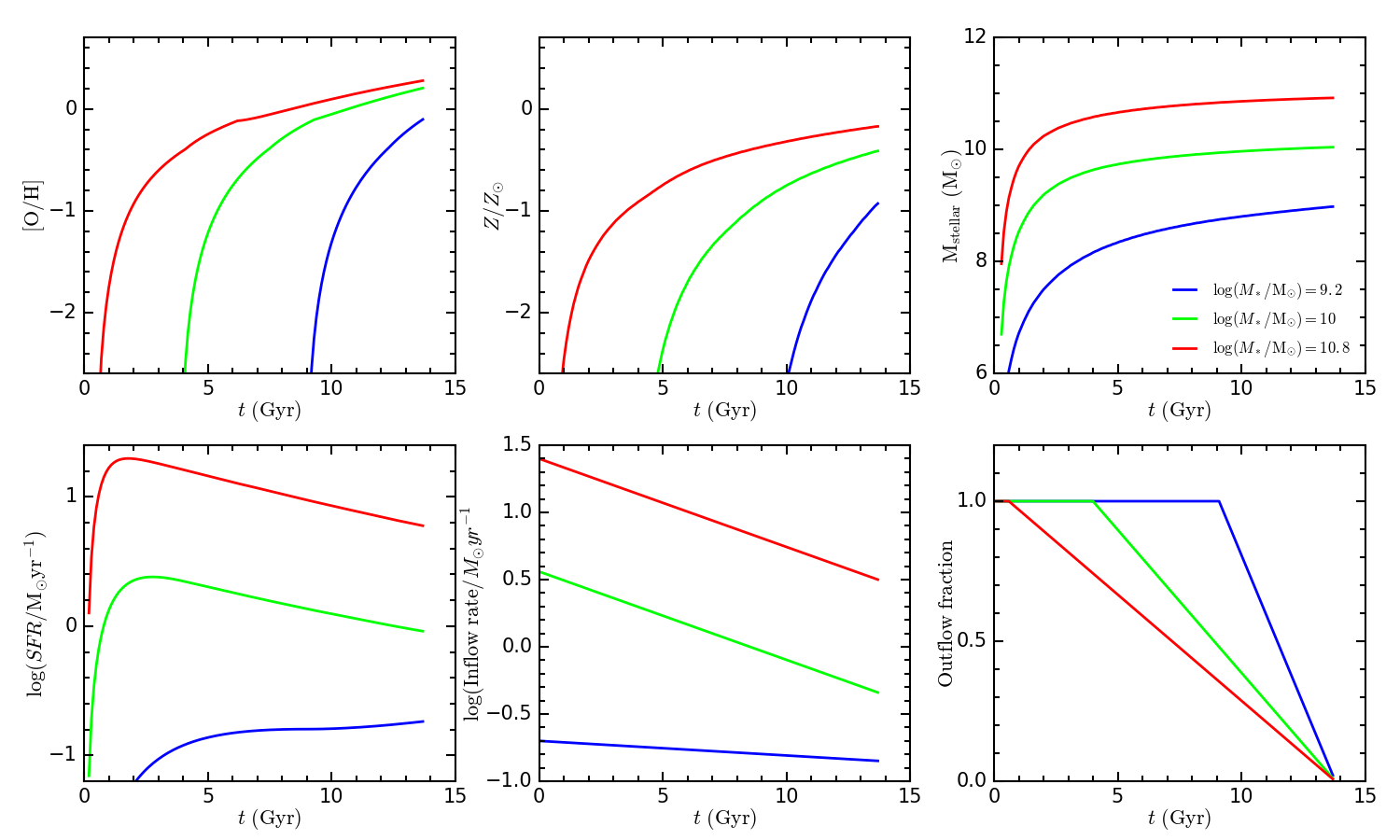}
\caption{Evolution of metallicity, stellar mass, SFR, inflow rate and metal outflow fraction as predicted by the variable metal outflow models in Figure 16. {The prediction for galaxies at different stellar masses is shown with different colours as indicated by the legend in the top right-hand panel.}}
\label{figure17}
\end{figure*}

\begin{table*}
\caption{Parameter ranges of the variable metal outflow model in Figure 16.}
\label{table3}
\centering
\begin{tabular}{l c c c c c c c c c c c c c c}
\hline\hline
$A_{\rm ks}$ & $A_{\rm ks,i}$ & $n_{\rm ks}$ & $n_{\rm ks,i}$ & $A_{\rm inf}$ & $\tau_{\rm inf}$ & $f_{\rm out,i}$$^{a}$ & $f_{\rm out,p}$$^{b}$& $t_{\rm out}$ & $\alpha1$ & $\alpha1_{\rm i}$ & $\alpha2$ & $\alpha2_{\rm i}$ & $t_{\rm delay}$ \\
  &	& & & ${\rm M_{\odot} yr}^{-1}$ & Gyr & & & Gyr & & & & & Gyr \\
\hline
0.50 & 0.50 & 1.5 & 1.5 & [0.20,25.11] & [40,6.6] & 100\% & 0 & [9,0.5] & 1.3 & 1.3 & 2.3 & 2.3 & 0 \\
\hline
Notes:\\
$a$: Initial outflow fraction.\\
$b$: Present outflow fraction. 
\end{tabular}\\  
\end{table*}

\subsubsection{Variable IMF model}
A model with time-dependent IMF slope is also able to match the metallicity observations
as discussed in \textsection3.2. Similar to the outflow model, we generate a series of models with a range of $A_{\rm inf}$
to reproduce the dynamic range in stellar mass of the galaxy sample. Other parameters are tuned as a function of 
$A_{\rm inf}$ to match the {MZR$_{\rm gas}$, MZR$_{\rm star}$, and mass-SFR relation}. 
Figure 18 and Figure 20 show the resulting models with varying {\sl initial} IMF slope at the low mass end $\alpha1_{\rm i}$ and at the high mass end $\alpha2_{\rm i}$, respectively. The parameter settings of these two variable IMF models are included in Table 4. 
Figure 19 and Figure 21 show the evolution of metallicity, stellar mass, SFR, inflow rate and outflow fraction as predicted by the variable IMF models. 

It can be seen that the models with mass-dependent initial IMF slope fit the {MZR$_{\rm gas}$ and the MZR$_{\rm star}$} equally well. 
In this variable IMF scenario, less-massive galaxies are assumed to have a {\em steeper} initial IMF slope, either at the low stellar mass end or at the high stellar mass end. The late-time IMF is fixed to the bimodal IMF with slopes at the low mass and at the high mass ends equal to the Kroupa IMF \citep{kroupa2001}. 
As we can see from Figure 10, the variation in the initial IMF slope changes the gas metallicity (even though much less than the stellar metallicity) comparable to the dynamic range of the observed gas metallicities in star-forming galaxies. Therefore, the mass-dependent initial IMF slope, if adopted to explain the {MZR$_{\rm star}$}, additionally {drives} the {MZR$_{\rm gas}$}. 

\subsubsection{Outflow vs IMF}
Observations at high redshift will be instrumental to distinguish between the two successful scenarios presented here. From the evolution of the galaxy properties in Figure 16 it can be seen that the variable outflow model predicts a mild evolution in zero point and slope of the {MZR$_{\rm gas}$} for the last several Gyrs with a much stronger evolution at early epochs. The variable IMF model, instead, predicts a much more dramatic evolution of the {MZR$_{\rm gas}$} at recent epochs (Figures~19 and 21). This difference in the cosmic evolution of the {MZR$_{\rm gas}$} between our two successful scenarios shows that the additional consideration of the redshift evolution 
of gas metallicity will shed further light on the actual origin of the {MZR$_{\rm gas}$}, which will be subject of a future paper.

\begin{figure*}
\centering
\includegraphics[width=17cm,viewport=0 10 700 640,clip]{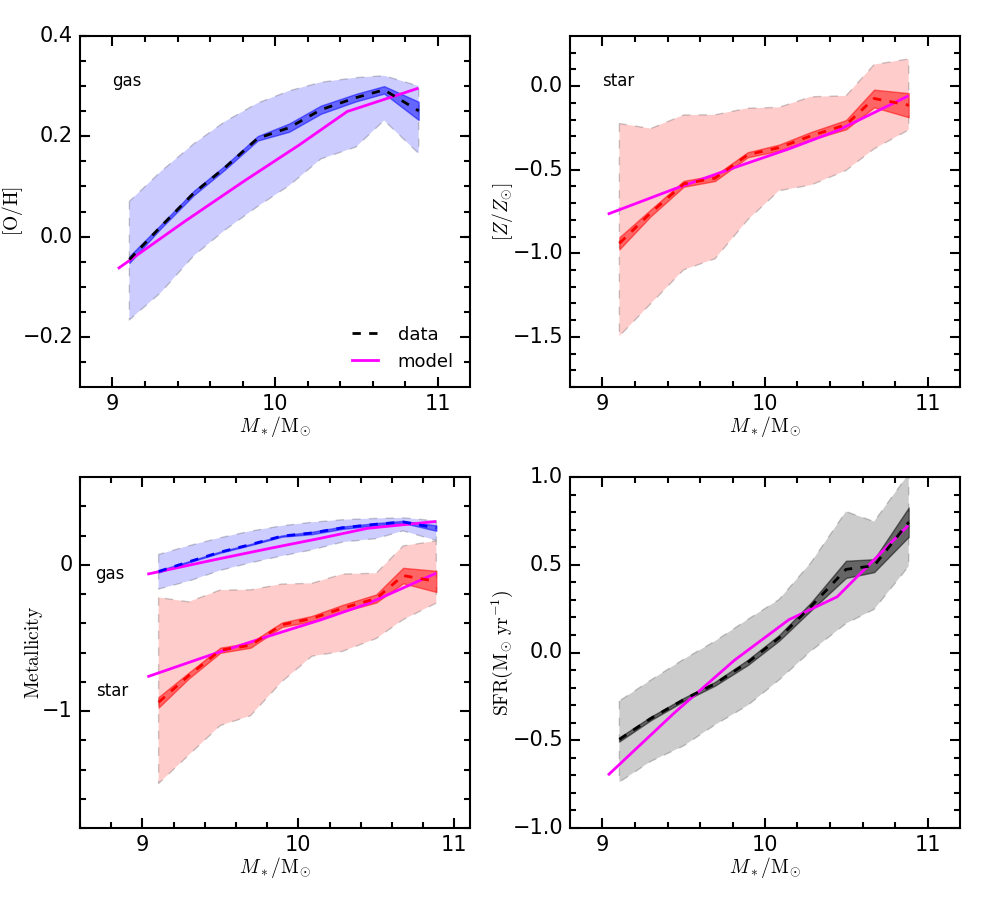}
\caption{Comparison between the observation and the finely-tuned variable IMF model for the {MZR$_{\rm gas}$, mass-weighted MZR$_{\rm star}$, and mass-SFR relation}. {Here we tune the initial IMF slope at low mass end.}}
\label{figure18}
\end{figure*}

\begin{figure*}
\centering
\includegraphics[width=17cm,viewport=10 10 1100 640,clip]{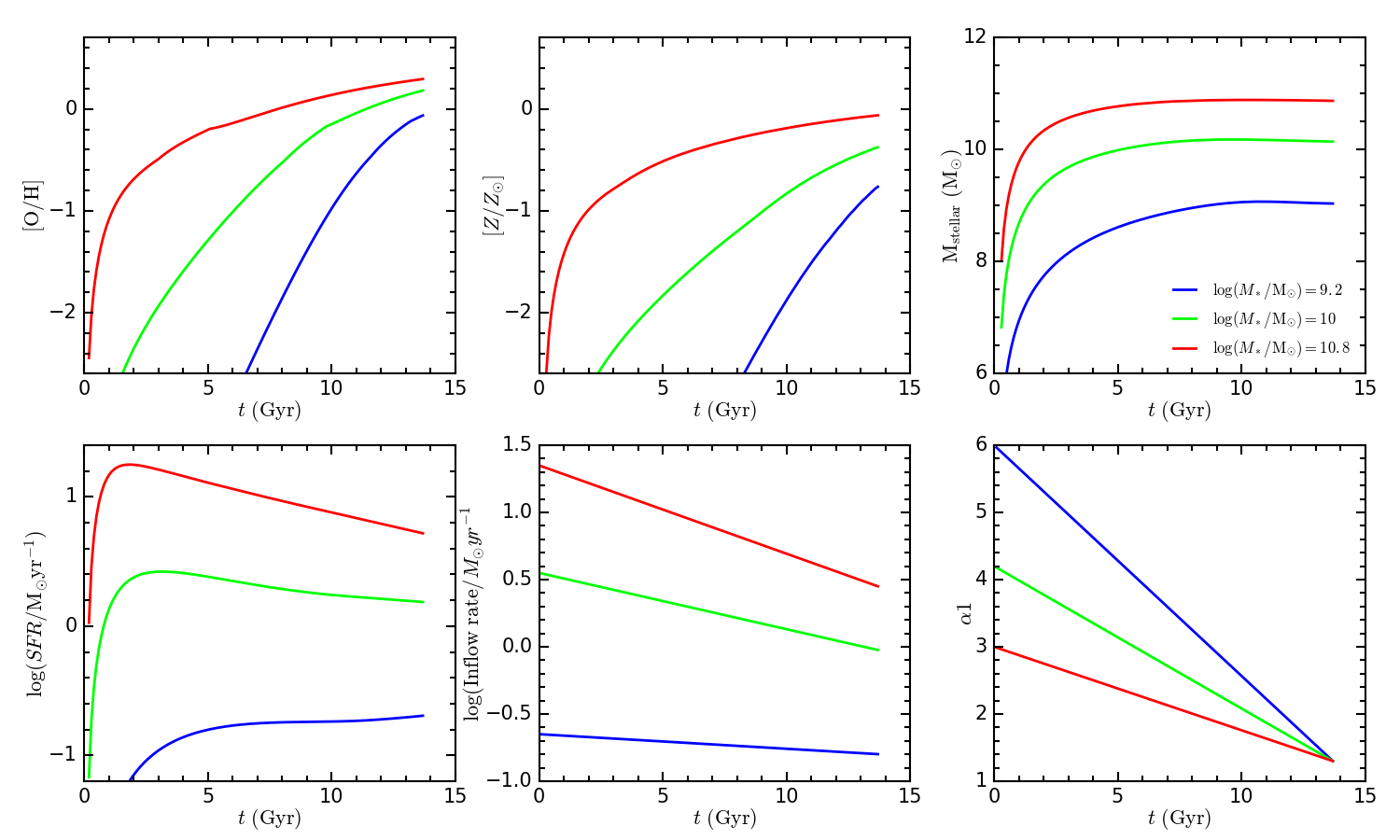}
\caption{Evolution of metallicity, stellar mass, SFR, inflow rate and outflow fraction as predicted by the variable IMF models of Figure 18.} 
\label{figure19}
\end{figure*}

\begin{figure*}
\centering
\includegraphics[width=17cm,viewport=0 10 700 640,clip]{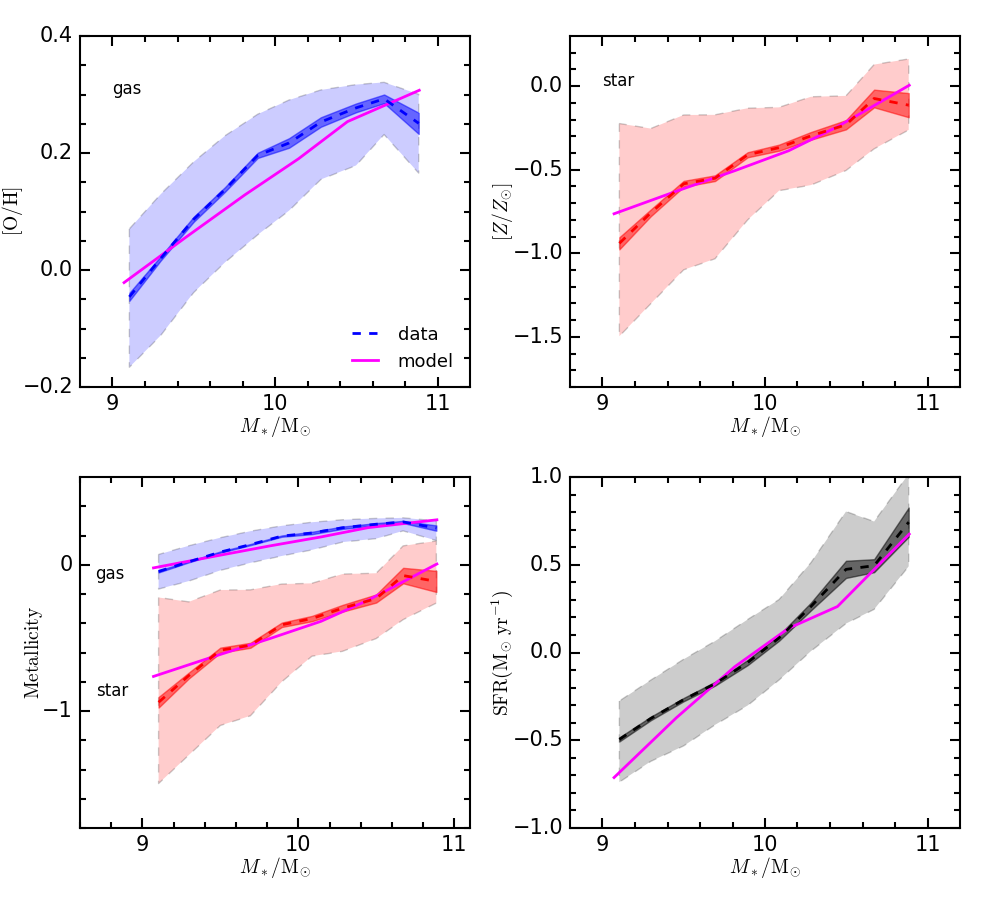}
\caption{Similar to Figure 18 but showing the IMF model with variable IMF slope at the high mass end.}
\label{figure20}
\end{figure*}

\begin{figure*}
\centering
\includegraphics[width=17cm,viewport=10 10 1100 640,clip]{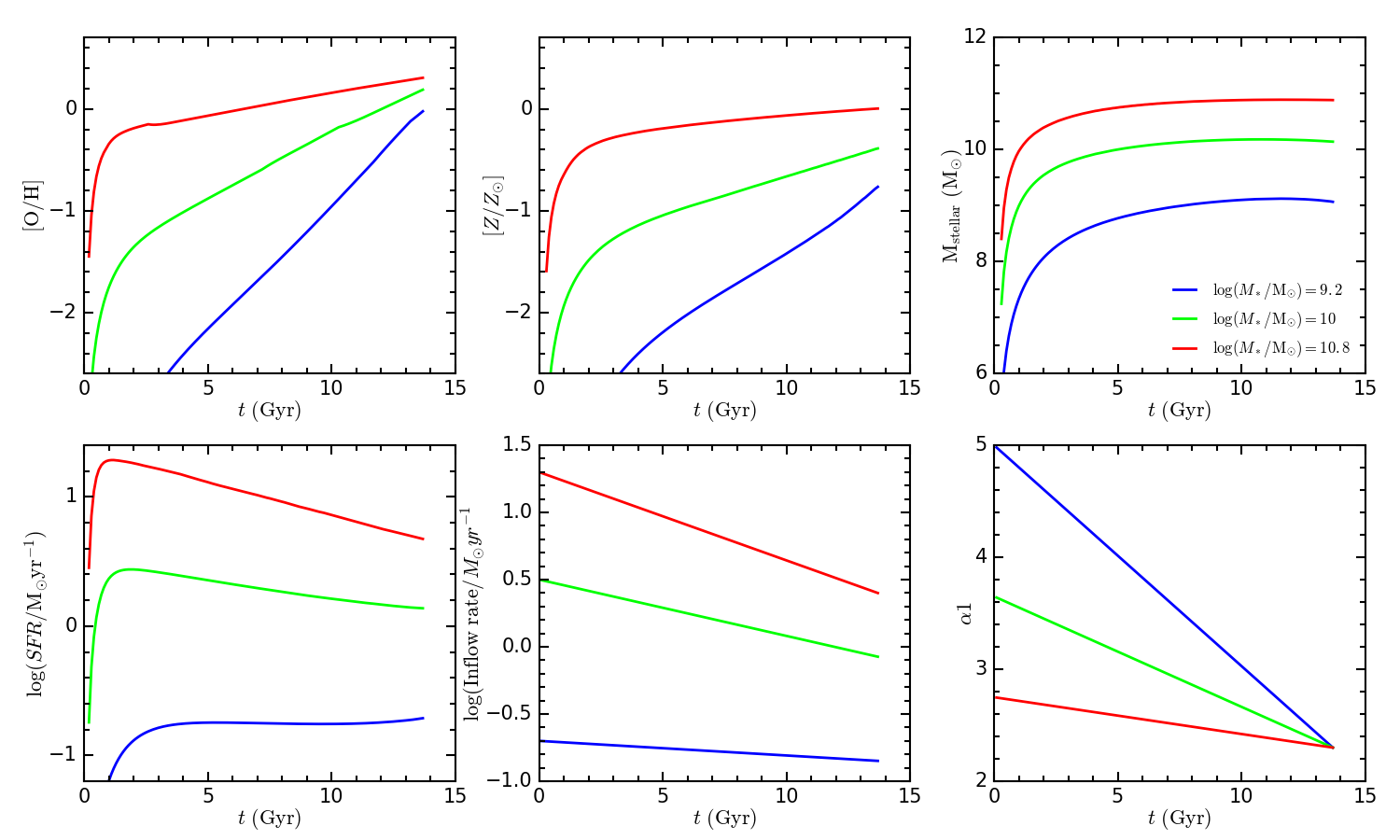}
\caption{Evolution of metallicity, stellar mass, SFR, inflow rate and outflow fraction as predicted by the variable IMF models in Figure 20.} 
\label{figure21}
\end{figure*}

\setlength{\tabcolsep}{4pt}

\begin{table*}
\caption{Parameter ranges of the variable IMF models of Figure 18 and Figure 20.}
\label{table4}
\centering
\begin{tabular}{l c c c c c c c c c c c c c c}
\hline\hline
models &	$A_{\rm ks}$ & $A_{\rm ks,i}$ & $n_{\rm ks,i}$ & $n_{\rm ks,i}$ & $A_{\rm inf}$ & $\tau_{\rm inf}$ & $f_{\rm out}$ & $t_{\rm out}$ & $\alpha1$ & $\alpha1_{\rm i}$ & $\alpha2$ & $\alpha2_{\rm i}$ & $t_{\rm delay}$ \\
& & & & & ${\rm M_{\odot} yr}^{-1}$ & Gyr & & Gyr & & & & & Gyr \\
\hline
$\alpha1$ & 0.5 & 0.5 & 1.5 & 1.5 & [0.22,22.39] & [40,6.6] & 0 & 0 & 1.3 & [6,3] & 2.3 & 2.3 & 0 \\
$\alpha2$ & 1.6 & 1.6 & 1.5 & 1.5 & [0.20,19.95] & [40,6.6] & 0 & 0 & 1.3 & 1.3 & 2.3 & [5,2.75] & 0 \\
\hline
\end{tabular}\\  
\end{table*}  

\setlength{\tabcolsep}{6pt}

\section{Discussion}
\subsection{Models in the literature}
Several flavours of chemical evolution models have been used in the literature to study the origin of the observed {MZRs}, (such as \citealt{thomas1998,calura2009,spitoni2017}), including combinations with cosmological semi-analytical models \citep{yates2012} or cosmological hydrodynamic simulations \citep{guo2016,rossi2017}. Most of the previous chemical evolution models focused on explaining either the gas metallicity or the stellar metallicity of galaxies. \citet{calura2009}, for instance, studied the {MZR$_{\rm gas}$} of local galaxies separated by morphological type using a full chemical evolution model. By reproducing the {MZR$_{\rm gas}$} for different galaxy types, the authors {interpret} this relation as a natural result of a mass dependent SFE, irrespective of galaxy morphological type. Less massive galaxies are less efficient in star formation and therefore more metal poor. This is consistent with the `model-SFE' models presented here {in section \textsection4 (Figure 14 and Figure 15).} 
However, it should be noted that a chemical evolution model with a mass-dependent SFE alone is not able to reproduce both the {MZR$_{\rm gas}$ and MZR$_{\rm star}$} simultaneously. 

\citet{spitoni2017} investigated the {MZR$_{\rm star}$} of star-forming and passive galaxies as derived by \citet{peng2015} using an analytical chemical evolution model. The authors obtained predictions for gas mass, total mass, and metallicity of a galaxy by solving a set of differential equations for these quantities. The observed {MZR$_{\rm star}$} is well reproduced by their analytical chemical evolution model adopting a mass-dependent outflow rate. The outflow rate is assumed to be proportional to the SFR, and the {proportionality} factor is assumed to be mass dependent. Hence in this model, the mass loading factor, i.e.\ outflow rate per SFR, is higher in less massive star-forming galaxies. In fact it is plausible that a galaxy with a higher outflow rate has a lower stellar metallicity. This trend can also be seen in Figure 9 of this paper where stellar metallicity is lower for the model with higher {\sl late-time} outflow fraction. However, we show that the higher late-time outflow fraction also leads to a lower gas metallicity comparable to the stellar metallicity. As we show in Figure 4, the observed gas metallicity in star-forming galaxies are much higher than their stellar metallicity, though. Therefore, we conclude that a mass-dependent outflow rate that is constant with time is not able to reproduce the observed gas and stellar metallicities simultaneously. 

In addition to studies focusing on gas or stellar metallicity separately, there are several attempts in the literature to reconcile these two metallicity phases within one single chemical evolution model. Based on the semi-analytical model, L-GALAXIES, combined with chemical evolution descriptions, \citet{yates2012} generated a sample of model galaxies that recovers the observed {MZR$_{\rm gas}$ and mass-SFR relation} \citep{mannucci2010,lopez2010}. However, the model overestimates the stellar metallicities, in particular at the low mass end \citep{yates2012}. In other words the model cannot reproduce the gas and stellar metallicties simultaneously, and the predicted {MZR$_{\rm star}$} is too shallow. \citet{rossi2017} find similar results using the high-resolution hydrodynamic simulation EAGLE. The model successfully reproduces the observed gas metallicity at all masses but again overestimates the stellar metallicity particularly of low-mass galaxies. It should be noted that the observed stellar metallicities used in the comparisons above are taken from \citet{gallazzi2005} and \citet{panter2008} where passive galaxies are also included. Considering the large difference in the stellar metallicity between star-forming and passive galaxies ($\sim 0.4$ dex at $10^{9.5} {\rm M_{\odot}}$; \citealt{peng2015}), the discrepancy between observation and predictions by these models would be larger when focusing on star forming galaxies. 

\subsection{Uncertainties in gas metallicity}
The key observable used in this work to constrain the chemical evolution model is the large difference between the gas and stellar metallicities. Therefore our results rely on how well the gas and stellar metallicities are measured. 

It is well known that gas metallicities derived by different strong emission line calibrations are not consistent and show large discrepancies (see \citealt{kewley2008} for a detailed discussion). The {MZR$_{\rm gas}$} obtained with different metallicity indicators, as a result, deviate from each other. To check whether our results depend on the choice of the gas metallicity indicator, we repeat our analysis (originally based on the theoretical R23 method) using gas metallicities obtained through the empirical N2 method.

{Figure A1} in appendix A shows the resulting {variable IMF} model that matches the {MZR$_{\rm gas}$, MZR$_{\rm star}$, and mass-SFR relation}. Since the empirical N2 method gives a much lower gas metallicity than the other theoretical methods, one of the following parameter settings is required based on the discussion in \textsection3.2\ to match the gas metallicity: an extremely low $A_{\rm ks}$ ($\sim$1 per cent of the original value of \citet{kennicutt1998}) or a strong late-time outflow fraction of $\sim 70$ per cent or a steep present IMF slope with $\alpha1\sim2.5$ or $\alpha2\sim2.8$. Here we choose the model with an outflow fraction of $\sim 70$ per cent. However, these parameters appear extreme and seem to lack observational evidence in the local universe where no clear signature of galactic outflow \citep{concas2017} or steep IMF \citep{bastian2010} is found in local star-forming galaxies. Hence generally we prefer the higher gas metallicities obtained from the theoretical R23 method. 

A consequence of the lower gas metallicities is that the difference between gas and stellar metallicities decreases. However, {as we show in section \textsection4, even adopting the gas metallicity derived through the N2 method, the chemical evolution model with normal parameter settings still fails to match the correct slopes of the {MZR$_{\rm gas}$ and MZR$_{\rm star}$} simultaneously.}
Hence, the general conclusion of this paper remains unchanged, and scenarios with either time-dependent metal outflow rates or time-dependent IMF slopes are still needed. The detail of the model parameters are affected, though. The absolute value of $a1_{\rm i}$ is slightly lower than in the variable IMF of Figure 20, for instance. Therefore, we conclude that although the absolute parameter values of our models will vary, the trends we find in variable outflow and variable IMF models are robust against the various gas metallicity calibrations.

\subsection{Uncertainties in stellar metallicity}

The intrinsic uncertainty in the observed stellar metallicity is mainly caused by the well-known degeneracy between the stellar metallicity, stellar age, and dust extinction. Despite the notable scatter in the {MZR$_{\rm star}$}, however, it should be noted that 
we should be more careful about any factors that could induce systematic offset in stellar metallicity. 

Nebular emission which originates from {ionized} gas around young and massive stars consists of not only line emission but also continuum emission. The strength of nebular continuum emission depends on the {ionization} radiation field and the metallicity of the gas.
In the star-forming galaxies, the spectra continuum is a combination of the continuum emission from stars and {ionized} gas. Therefore, it is expected that the intrinsic equivalent width (EW) of the absorption lines will be higher than the observed value. Since the stellar metallicity derived from absorption lines is positively correlated with the line strength, the underestimation in the absorption lines will result in a underestimation of stellar metallicity \citep{alfonso1996}.
  
Recently, \citet{byler2016} {analyzed} the nebular emission under different environments in detail using stellar population synthesis models. From Figure~12 in their paper, we can see that, regardless of the gas metallicity, the nebular continuum emission is negligible (less than 1\% by eye) compared to the stellar continuum emission when the age of the single stellar population is above 0.01 Gyr. The typical age of the galaxies in our sample with a median of 4.98 Gyr is much higher than this limit. Moreover, we use a full spectra fitting algorithm which {utilizes} not only the absorption lines but also the continuum shape to constrain the stellar metallicity. Therefore we conclude that the stellar metallicity used in this work is not affected by the nebular continuum contamination and our results are robust.   

\subsection{Outflows without AGB winds}
We assume the galactic outflow of metals to consist of stellar ejecta from SN-II, SN-Ia, and AGB winds. Supernovae release
energies large enough to expel the ejecta from a galaxy. AGB winds, instead, are less energetic, and will not be able to drive galactic mass loss by themselves. Hence our model implies that AGB ejecta are dragged along with the galactic outflow generated by supernovae. To test whether this assumption affects our analysis, we also explore chemical evolution models where AGB winds are not included the outflow.
 
{Figure B1} in Appendix B shows the evolution of gas and stellar metallicities and the SFR as predicted by models with varying metal outflow fraction {\em excluding AGB winds}. It turns out that the {final} gas and stellar metallicities do not change significantly for scenarios with {both} time-constant {and time-dependent} outflow rates. This is due to the fact that the metal production of a galaxy is dominated by enrichment through SN-II supernovae. If AGB winds are excluded from the galactic outflow, the final gas metallicity remains unchanged while the final stellar metallicity increases only slightly by $\sim 0.1$ dex. The conclusions of this paper remain unchanged.

\section{Conclusions}
In this work we investigate the gas and the mass-weighted stellar metallicities of local star-forming galaxies. We select a sample of local star forming galaxies from SDSS DR12 and measure gas metallicities using the empirical N2 method and the theoretical R23 method, as well as mass- and light-weighted stellar metallicities derived from full spectral fitting. The latter is done using the code FIREFLY (Wilkinson et al 2015, in prep) which fits a full array of single burst stellar population models (from \citealt{maraston2011}) and retains individual and linear combinations of solutions, without adopting any particular prior in the star formation history. Metallicity is output as both light-weighted and mass-weighted.   
It turns out that the mass-weighted stellar metallicities of star-forming galaxies are much lower than their gas metallicities.
The difference reaches $\sim0.8$ for galaxies {with} $10^{9} {\rm M_{\odot}}$ when using gas metallicities obtained through the theoretical R23 method, and $\sim0.4$ dex when using gas metallicities obtained through the empirical N2 method. 
Even more interestingly, this discrepancy between gas and stellar metallicities is mass dependent with larger offsets in 
lower mass galaxies. In other words, the {MZR$_{\rm gas}$} is much flatter than the {MZR$_{\rm star}$}.
Since gas metallicity represents the current metal abundance of a galaxy while stellar metallicity carries information about metal enrichment at early epochs, a much lower stellar metallicity implies significant metallicity evolution with much lower metal abundances at early times.

To investigate the origin of the {MZR$_{\rm gas}$ and MZR$_{\rm star}$} of star-forming galaxies, we construct a simple galactic 
chemical evolution model. There are eight free parameters in the model: the coefficient and index of the KS star formation law, $A_{\rm ks}$,
$a_{\rm ks}$, the initial inflow strength, $A_{\rm inf}$ with the declining time-scale, $\tau_{\rm inf}$, the outflow fraction, $f_{\rm out}$, the IMF slope at the low stellar mass end, $\alpha1$, and at the high stellar mass end, $\alpha2$, and finally the time-delay in mixing of stellar ejecta with the ISM, $t_{\rm delay}$. For each parameter, we explore the parameter space to
investigate potential degeneracies, and the ability to reproduce the {MZR$_{\rm gas}$ and MZR$_{\rm star}$}. As far as the gas metallicity is concerned, the degeneracy is strong as most of these parameters affect the gas metallicity significantly. 
The observed stellar metallicity, instead, cannot be matched easily by a simple variation of these parameters. To match the significant suppression of metal enrichment at early times as implied by the data, we explore chemical evolution models imposing a time dependence of these parameters. We find that models with a strong metal outflow or a steep IMF slope at early times in the evolution of a galaxy succeed in explaining the observed large difference between the gas and stellar metallicities. 

To directly compare the model with observations, we generate a series of models with a range of inflow strengths $A_{\rm inf}$ 
matching the dynamic range of stellar mass of the galaxy sample. We then modify other parameters at a given $A_{\rm inf}$ 
to match the SFR, the gas and the stellar metallicities at a given galaxy mass. We find that models with a strong time dependence of either the metal outflow {fraction} or the IMF slope match the observations very well. Either enhanced metal outflows or steeper IMF slopes at early times, particularly in low-mass galaxies, are required to match the observed difference in gas and stellar metallicities. This is the first time that the 
{MZR$_{\rm gas}$ and MZR$_{\rm star}$} of galaxies are reproduced simultaneously by a chemical evolution model.

In the `best-fit' outflow model, low mass galaxies tend to have a later outflow transition time $t_{\rm out}$ and hence stronger metal outflows {on average}. This is astrophysically plausible since the gravitational potential well is shallower for galaxies with lower masses and at early epochs. In the `best-fit' IMF slope model, low mass galaxies
are required to have a steeper initial IMF slope with $\alpha1_{\rm i}\sim4.3$ or $\alpha2_{\rm i}\sim4.3$ at $10^9{\rm M_{\odot}}$ at early epochs to efficiently suppress the metal enrichment.
It should be noted that in addition to {the MZR$_{\rm gas}$ and MZR$_{\rm star}$, the mass-SFR} relation is also reproduced by {the two} scenarios.
{In these two scenarios, the stellar metallicity of galaxies are driven by the average metal outflow fraction or average IMF slope.  
As a result, the MZR$_{\rm star}$ is regulated by a mass-dependent average metal outflow fraction or average IMF slope. 
In contrast, the gas metallicity is driven by more recent evolutionary processes. Any mass dependence of these processes, such as a mass-dependent recent metal outflow fraction, will result in the observed MZR$_{\rm gas}$.}

As the gas metallicities obtained from the empirical N2 method are considerably lower than the ones obtained from the theoretical R23 method, the discrepancy between gas and stellar metallicities reduces and it is easier for the models in principle to match gas and stellar metallicities simultaneously. However, as we discuss in the present paper, the slopes of the two relationships are still different, and models with time-dependent metal outflow or IMF slope as discussed are still required, even if the relatively low gas metallicities from empirical calibrations are adopted. On top of this, our chemical evolution model tends to over-predict gas metallicities when compared to N2 gas metallicities. To still fit the observations, extreme parameter settings such as unrealistic low star formation efficiencies or {strong present outflow or steep present IMF slope} need to be adopted, which can be taken as an indication in favour of the higher gas metallicities obtained from the theoretical calibrations like the R23 method.
Clearly, further theoretical work and observational constraints are needed to break the degeneracy in the model parameters. The inclusion of observational constraints both spatially resolved and as a function of cosmic time will be subject of future papers.

\section*{Acknowledgements}
We are grateful to the referee's useful and constructive comments and suggestions which improve the quality of this paper.

\appendix
\section{Gas metallicity by N2 method}
Figure A1 shows the model tuned to match the {MZR$_{\rm gas}$ by the N2 method, MZR$_{\rm star}$, and mass-SFR relation}. For an example, we only show one effective model with a variable IMF slope at high mass end. The parameter ranges are listed in {Table A1}.

\begin{figure*}
\centering
\includegraphics[width=17cm,viewport=0 10 700 640,clip]{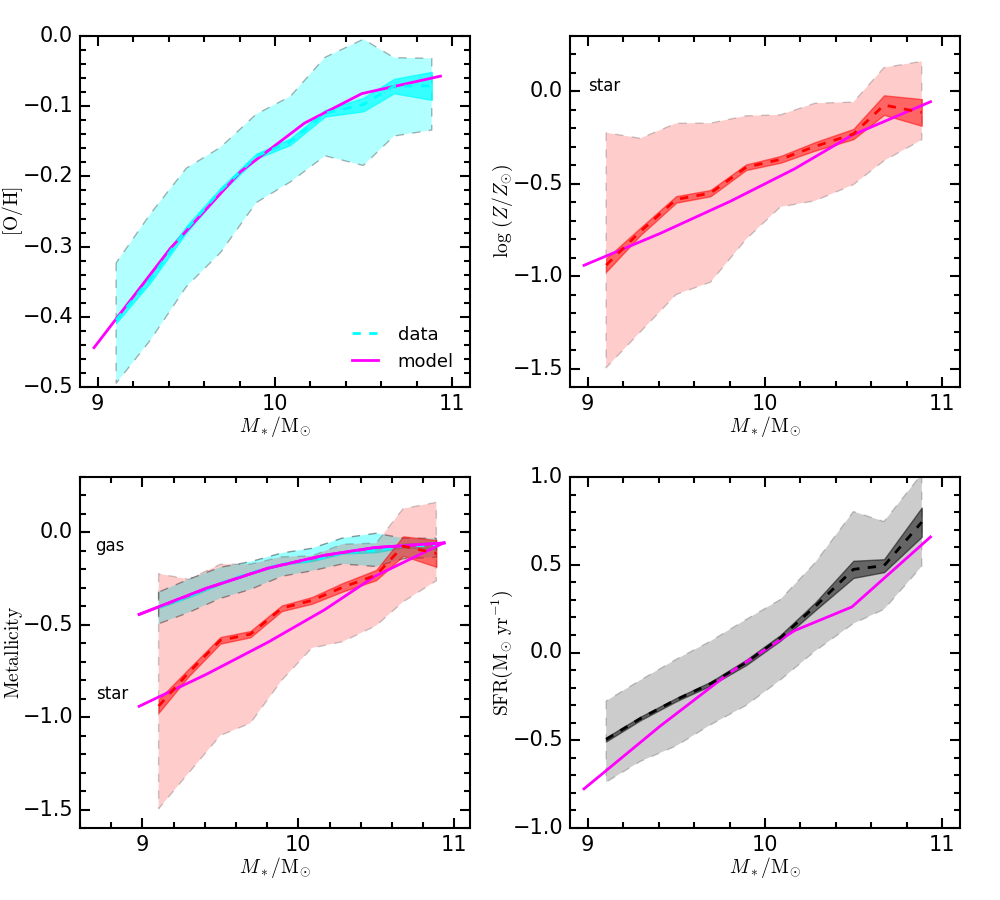}
\caption{Similar to Figure 20 but showing the {MZR$_{\rm gas}$} by the N2 method and the finely-tuned variable IMF model.}
\label{figurea1}
\end{figure*}

\begin{table*}
	\caption{Parameter ranges of the variable IMF model that matches the {MZR$_{\rm gas}$ with} the N2 method in Figure A1.}
	\label{table5}
	\centering
	\begin{tabular}{l c c c c c c c c c c c c c c}
		\hline\hline
		$A_{\rm ks}$ & $A_{\rm ks,i}$ & $n_{\rm ks}$ & $n_{\rm ks,i}$ & $A_{\rm inf}$ & $\tau_{\rm inf}$ & $f_{\rm out}$ & $t_{\rm out}$ & $\alpha1$ & $\alpha1_{\rm i}$ & $\alpha2$ & $\alpha2_{\rm i}$ & $t_{\rm delay}$ \\
		& & & & ${\rm M_{\odot} yr}^{-1}$ & Gyr & & Gyr & & & & & Gyr \\
		\hline
		0.28 & 0.28 & 1.5 & 1.5 & [0.20,25.12] & [40,6.6] & 0.60 & 0 & 1.3 & 1.3 & [3.5,2.1] & 2.3 & 0 \\
		\hline
	\end{tabular}\\  
\end{table*} 

\section{Outflow without AGB winds}
Figure B1 shows the evolution of metallicities and SFR predicted by the model with various outflow fraction in which the AGB winds are excluded.

\begin{figure*}
\centering
\includegraphics[width=18cm]{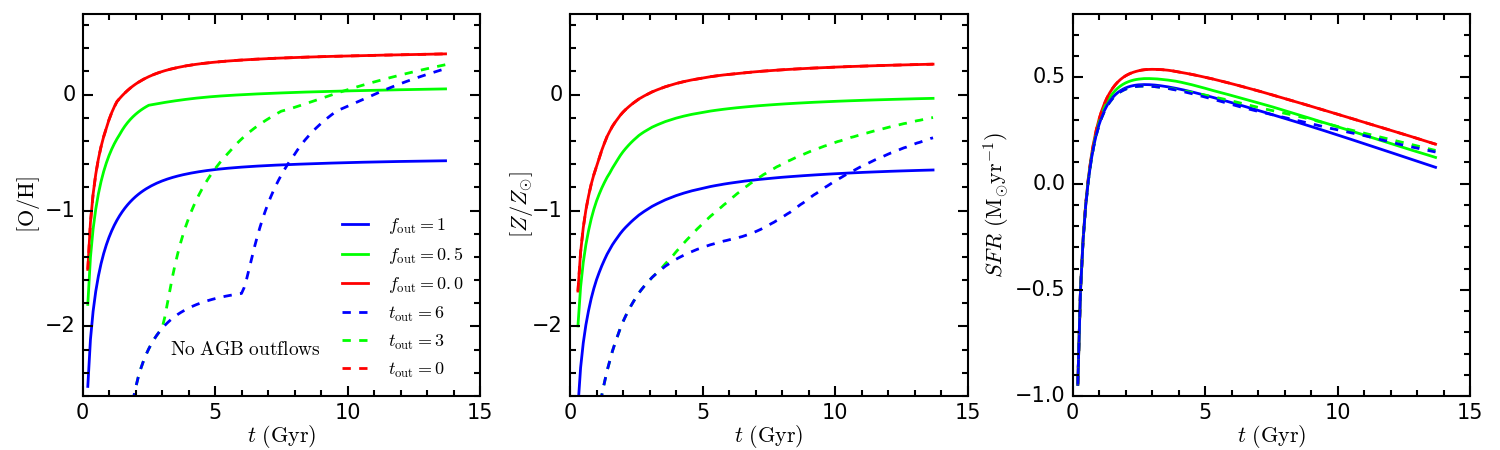}
\caption{Predicted evolution of gas metallicity, stellar metallicity, and SFR by the chemical evolution model with different outflow fraction $f_{\rm out}$ and transition time $t_{\rm out}$. Here the mass ejecta by AGB winds are excluded from the galactic outflow and hence remaining in the galaxy. }
\label{Figureb1}
\end{figure*}

\end{document}